\newcommand{\diracslash}[1]{#1\llap{/\kern2pt}}

\newcommand{\be}{\begin{equation}}
\newcommand{\ee}{\end{equation}}
\newcommand{\bea}{\begin{eqnarray}}
\newcommand{\eea}{\end{eqnarray}}
\newcommand{\ba}[1]{\begin{array}{#1}}
\newcommand{\ea}{\end{array}}

\newcommand{\bt}{\begin{tabular}}
\newcommand{\et}{\end{tabular}}

\newcommand{\beas}{\begin{eqnarray*}}
\newcommand{\eeas}{\end{eqnarray*}}

\documentclass[preprint,prd,aps,floats,nofootinbib,showpacs,floatfix]{revtex4}
\usepackage{graphicx}
\begin{document}

\title{D-mesons in isospin asymmetric strange hadronic matter }
\author{Arvind Kumar}
\email{iitd.arvind@gmail.com}
\affiliation{Department of Physics, Indian Institute of Technology, Delhi,
Hauz Khas, New Delhi -- 110 016, India}

\author{Amruta Mishra}
\email{amruta@physics.iitd.ac.in,mishra@th.physik.uni-frankfurt.de}
\affiliation{Department of Physics, Indian Institute of Technology, Delhi,
Hauz Khas, New Delhi -- 110 016, India}

\begin{abstract}
We study the in-medium properties of $D$ and $\bar{D}$ mesons in isospin
asymmetric hyperonic matter arising due to their interactions with
the light hadrons. The interactions of $D$ and $\bar{D}$ mesons with 
these light hadrons are derived by generalizing the chiral SU(3) model
used for the study of hyperonic matter to SU(4). The nucleons, the scalar
isoscalar meson, $\sigma$ and the scalar-isovector meson, $\delta$ as modified 
in the strange hadronic matter, modify the masses of $D$ and $\bar{D}$ mesons. 
It is found that as compared to the $\bar{D}$ mesons, the $D$ meson 
properties are more sensitive to the isospin asymmetry at high densities. 
The effects of strangeness in the medium on the properties of $D$ and 
$\bar{D}$ mesons are studied in the present investigation. The $D$ mesons 
($D^0$,$D^+$) are found to undergo larger medium modifications as compared 
to $\bar{D}$ mesons ($\bar {D^0}$, $D^-$) with the strangeness fraction, 
$f_s$ and these modifications are observed to be more appreciable 
at high densities. The present study of the in-medium properties 
of $D$ and $\bar{D}$ mesons will be of relevance for the experiments 
in the future Facility for Antiproton and Ion Research, GSI, where 
the baryonic matter at high densities will be produced. The isospin 
asymmetric effects in the doublet $D = (D^{0}, D^{+})$ in the strange 
hadronic matter should show in observables like their production and 
flow in asymmetric heavy-ion collisions as well as in $J/\psi$ suppression. 

\end{abstract}
\pacs{24.10.Cn; 13.75.Jz; 25.75.-q}
\maketitle

\def\bfm#1{\mbox{\boldmath $#1$}}

\section{Introduction}
Since long time the topic of study of the in-medium properties of hadrons 
has drawn considerable interest of particle physicst because of its 
importance in understanding the strong interaction physics and having 
relevance in heavy-ion collision experiments as well as in nuclear
astrophysics. There have been extensive experimental efforts to study 
the in-medium properties of hadrons through experimental observables 
like production and propagation of the hadrons in hot and dense hadronic 
medium. The observed enhanced dilepton spectra \cite{ceres,helios,dls} 
could be a signature of in medium  vector meson mass reduction 
\cite{Brat1,CB99,vecmass,dilepton,liko} and for the formation of 
thermalized quark gluon plasma \cite{dk1}. The production of photons 
\cite{dk2} arising from the heavy ion collision experiments is a promising 
observable which probes matter resulting from the high energy nuclear 
collisions. The hard photons production is a promising source that 
can provide information about the thermodynamical state of the plasma 
produced in heavy ion collision experiments \cite{bhatt1}. Similarly 
kaons and antikaons properties have been studied experimentally
by KaoS collaboration and the production of kaons and antikaons 
in the heavy-ion collisions and their collective flow are directly 
related to the medium modification of their spectral functions 
\cite{CB99,cmko,lix,Li2001,K5,K6,K4,kaosnew}. 
At the Compressed baryonic matter (CBM) experiment at the future
facility at GSI, one expects to produce matter at high densities
\cite{gsi} and the in-medium properties of strange and charm mesons
are some of the important topics planned to be investigated extensively. 
Therefore, the topic of the study of charm mesons in dense hadronic matter
has also gotten considerable interest in the recent past.
The $D$ and $\bar{D}$ mesons are modified due to 
interaction of the light quark (u,d) or antiquark present in them
with the nuclear medium. The experimental signatures for the mass
modifications of $D$ and $\bar D$ mesons can be their production ratio 
as well as $J/\psi$ suppression in the hadronic medium
\cite{NA501,NA50e,NA502}. In heavy-ion collision 
experiments of much higher energies, e.g., in RHIC or LHC, 
it is suggested that the $J/\psi$ suppression
is because of the formation of quark-gluon plasma (QGP) \cite{blaiz,satz}. 
However, in Ref. \cite{zhang,brat5,elena} it is observed that the effect 
of hadron absorption of $J/\psi$ is not negligible. 
The $J/\psi$ suppression can also be 
due to comover interaction as suggested in Ref. \cite{capella}.
An important difference between $J/\psi$ suppression 
pattern in comovers interaction model and in
a deconfining scenario is that in the former case
the anomalous suppression sets in smoothly from 
peripheral to central collisions rather than in 
a sudden way when the deconfining threshold is reached \cite{capella}. 
In Ref. \cite{vog}, it was reported that the charmonium
suppression observed in Pb + Pb collisions 
of NA50 experiment cannot be explained
by nucleon absorption only, but need some additional 
density dependent suppression mechanism.
It was suggested in these studies that the comover scattering can
explain the additional suppression of charmonium. 
The $J/\psi$ suppression in nuclear collision
at SPS energies has been studied in covariant 
transport approach HSD in Ref. \cite{cassing}.
The calculations show that the absorption of $J/\psi$'s by both
nucleons and produced mesons can explain reasonably not
only the total $J/\psi$ cross section but also the transverse 
energy dependence of $J/\psi$ suppression measured in 
both proton-nucleus and nucleus-nucleus collisions.
Due to the reduction 
in the masses of $D$ and $\bar{D}$ mesons in the medium it is a possibility that
excited charmonium states can decay to $D\bar{D}$ pairs \cite{brat6} 
instead of decaying to lowest charmonium state $J/\psi$. 
Actually higher charmonium states are considered as the major 
source of $J/\psi$ \cite{pAdata}. Even at certain higher densities 
it can become 
a possibility that $J/\psi$ itself will decay to $D\bar{D}$ pairs.
So this can be an explanation of the observed $J/\psi$ suppression 
by NA50 collaboration at $158$ GeV/nucleon in the Pb-Pb collisions 
\cite{blaiz}. 
The excited states of charmonium also undergo mass drop in the 
nuclear medium \cite{leeko}.
The modifications of the in-medium masses of $D$ mesons is larger
than the $J/\psi$ mass modification \cite{haya1,friman}. This is 
because $J/\psi$ is
made of heavy quarks and these interact with the nuclear medium through 
gluon condensates. The change in gluon condensates with the nuclear density 
is very small and hence leads to only a small modification of the
$J/\psi$ mass.

The in-medium modifications of $D$ and $\bar{D}$ mesons have been studied in 
the literature using QCD sum rule approach, 
quark meson coupling (QMC) model and coupled 
channel approach. In the QCD sum rule approach, it 
is suggested that the light quark or antiquark of $D$ mesons interact with
the light quark condensate leading to the medium modification 
of $D (\bar{D})$ meson masses \cite{arata,qcdsum08}.
In the QMC model the scalar $\sigma$ meson couples
to the confined light quark (u,d) in the nucleon thus giving a drop of 
the nucleon mass in the medium . The D -meson properties within the 
QMC model have been studied in Ref. \cite{qmc}. The drop in the mass 
of $D$ mesons arises due to the interaction with the nuclear medium 
and the mass drop of the $D$ mesons observed in the QMC model turns 
out to be similar to that calculated in the QCD sum rule approach.
Charmed baryonic resonances have recieved a lot of attentions due to discovery
of quite a few new states by CELO, Belle, and BABAR Collaborations 
\cite{artuso,mizuk1,chistov,aubert1,aubert2,mizuk2}.
Whether these resonance states have $qqq$ structure or generated dynamcially
via meson baryon scattering processes is a matter of great interest 
\cite{ltolosdstar}. 
The study of the properties of $D$ mesons in the coupled channel approach 
using separable potential leads to the generation of resonance 
$\Lambda(2593)$ in I = 0 channel \cite{ltolos} analogous to
$\Lambda (1405)$ in the coupled channel approach for the $\bar K N$ interaction
\cite{kbarn}. The study of the spectral density of $D$ mesons at finite 
temperatures and densities, considering modifications of the nucleons
in the medium, indicate a dominant increase in the width of 
the $D$-meson whereas there is a very small change in the $D$-meson mass 
in the medium \cite{ljhs}. However, these calculations
\cite{ltolos,ljhs}, assume the interaction to be SU(3) symmetric 
in u,d,c quarks and ignore channels with charmed hadrons with strangeness.
A coupled channel approach for the study of D-mesons has been developed based
on SU(4) symmetry \cite{HL} to construct the effective interaction between 
pseudoscalar mesons in a 16-plet with baryons in 20-plet representation
through exchange of vector mesons and with KSFR condition \cite{KSFR}.
This model \cite{HL} has been modified in aspects like regularization method
and has been used to study DN interactions in Ref. \cite{mizutani6}. 
This reproduces the resonance  $\Lambda_c (2593)$ in the I=0 channel and in
addition generates another resonance in the I=1 channel at around 2770 MeV.
These calculations have been generalized to finite temperatures 
\cite{mizutani8} 
accounting for the in-medium modifications of the nucleons in a Walecka type
$\sigma-\omega$ model, to study the $D$ and $\bar D$ properties \cite{MK} 
in the hot and dense hadronic matter. At the nuclear matter density and for
zero temperature, these resonances ($\Lambda_c (2593)$ and $\Sigma_c (2770)$)
are generated $45$ MeV and $40$ MeV below their free space positions. 
However at finite temperature, e.g., at $T = 100$ MeV resonance positions 
shift to $2579$ MeV and $2767$ MeV for $\Lambda_{c}$  ($I = 0$) and 
$\Sigma_{c}$ ($I = 1$) respectively. Thus at finite temperature resonances 
are seen to move closer to their free space values. This is  
because of the reduction of pauli blocking factor arising due to the 
fact that fermi surface is smeared out with temperature. For $\bar{D}$ 
mesons in coupled channel approach a small repulsive mass shift is obtained. 
This will rule out of any possibility of charmed mesic nuclei \cite{mizutani8}
suggested in the QMC model \cite{qmc}. 
But as we shall see in our investigation, we obtain a small attractive mass 
shift for $\bar{D}$ mesons which can give rise to the possibility
of the formation of charmed mesic nuclei.
Using heavy quark symmetry, the $D^{0}$-nucleus bound states have been 
studied in Ref. \cite{garcia1} using
$D$ meson selfenergy calculated in the nuclear medium \cite{ltolosdstar}.
In heavy quark symmetry, the pseudoscalar $D$ meson and the vector meson, 
$D^{*}$ 
are treated on equal footing and this leads to the generation of a broad
spectrum of new resonant meson-baryon states in the charm one and strangeness zero \cite{garcia2}
and the exotic charm minus one \cite{gamer} sectors.
The inclusion of vector mesons is the keystone for
obtaining an attractive D-nucleus interaction that leads to the
existence of $D^{0}$ nucleus bound states, as compared previous studies
based on SU(4) flavor symmetry \cite{garcia1}.
The study of $D$ meson 
self-energy in the nuclear matter is also helpful in understanding
the properties of the charm and the hidden charm resonances in the 
nuclear matter \cite{tolosra}. In coupled channel 
approach the charmed resonance $D_{s0}(2317)$ mainly couples to $DK$ system, 
while the $D_{0}(2400)$ couples to $D\pi$ and $D_{s}\bar{K}$. The hidden 
charm resonance couples mostly to $D\bar{D}$. Therefore any modification 
of $D$ meson properties in the nuclear medium will affect the properties 
of these resonances. 

In the present investigation we study the in-medium modifications 
of the $D$ and $\bar{D}$ mesons in isospin asymmetric strange hadronic 
matter. The medium modifications of $D$ and $\bar{D}$ 
mesons are due to their interactions with the nucleons, the 
scalar isoscalar meson $\sigma$ and the scalar isovector meson $\delta$,
which are modified in the hyperonic matter. The medium modifications 
of the light hadrons (nucleons and scalar mesons) are described by 
using a chiral $SU(3)$ model \cite{paper3}. The model has been used 
to study finite nuclei, 
the nuclear matter properties, the in-medium properties of the vector mesons 
\cite{hartree,kristof1} as well as to investigate the optical potentials 
of kaons and antikaons in nuclear matter \cite{kmeson,isoamss} and in 
hyperonic matter in \cite{isoamss2}. For the study of the properties of $D$ 
mesons in isospin-asymmetric strange hadronic matter, the chiral 
SU(3) model is generalized to $SU(4)$ flavor symmetry to obtain the 
interactions of $D$ and $\bar{D}$ mesons with the light hadrons.
Since the chiral symmetry is explicitly broken for the SU(4) case
due to the large charm quark mass, we use the SU(4) symmetry here
only to obtain the interactions of the $D$ and $\bar D$ mesons with
the light hadron sector, but use the observed values of the heavy
hadron masses and empirical values of the decay constants. This 
has been in line with the philosophy followed in Ref. \cite{liukolin}
where charmonium absorption in nuclear matter was studied using 
the SU(4) model to obtain the relevant interactions. However,
the values of the heavy hadron masses and the coupling constants 
in Ref. \cite{liukolin}, were taken as the empirical values 
or as calculated from other theoretical models. The coupling
constants were derived by using the relations from SU(4) symmetry, 
if neither the empirical values nor values calculated from other 
theoretical models were available \cite{liukolin}. The 
$D$ meson properties in symmetric hot nuclear matter using the SU(4) 
model have been studied in Ref. \cite{amdmeson} and for the 
case of asymmetric nuclear matter at zero temperature \cite{amarind}
and at finite temperature in Ref. \cite{amarvind}. The charmonium mass 
in hot asymmetric nuclear matter hass been studied within present chiral 
model in Ref. \cite{amarvind, charmmass2}. The mass modifications of 
charmonium states arise due to the interaction with the gluon condensates
of QCD, simulated through the scalar dilaton field introduced to incorporate 
the broken scale invariance of QCD within effective chiral model 
Ref. \cite{amarvind}.

Within the SU(4) model considered in the present investigation, 
the $D(\bar D)$ energies are modified due to a vectorial Weinberg-Tomozawa, 
scalar exchange terms as well as range terms
\cite{isoamss,isoamss2}.
The isospin asymmetric effects among $D^0$ and $D^+$ in the doublet,
D$\equiv (D^0,D^+)$ as well as between $\bar {D^0}$ and $D^-$ in the
doublet, $\bar D \equiv (\bar {D^0},D^-)$ arise due to the scalar-isovector
$\delta$ meson, due to asymmetric contributions in the Weinberg-Tomozawa
term, as well as in the range term \cite{isoamss}. 
    
The outline of the paper is as follows : in section II, we give a brief 
introduction to the effective chiral $SU(3)$ model used to study 
the isospin asymmetric strange hadronic matter, and
its extension to the $SU(4)$ model to derive the interactions
of the charmed mesons with the light hadrons. In section III, we present 
the dispersion relations for the $D$ and $\bar{D}$ mesons which are solved 
to obtain their optical potentials in the dense strange hyperonic matter.
The effects of isospin asymmetry, strangeness fraction of the hadronic medium
on the optical potentials of the $D$ and $\bar D$ mesons are 
investigated. 
Section IV contains the results obtained for the medium modifications
of the $D$ and $\bar D$ mesons in the hyperonic matter and finally,
in section V, we summarize the findings of the present investigation 
and discuss possible outlook. 

\section{The hadronic chiral $SU(3) \times SU(3)$ model }

We use a chiral $SU(3)$ model for the study of the light hadrons in the 
present investigation \cite{paper3}. The model is based on nonlinear 
realization of chiral symmetry \cite{weinberg,coleman,bardeen} and 
broken scale invariance \cite{paper3,hartree,kristof1}. 
The effective hadronic chiral Lagrangian contains the following terms
\begin{equation}
{\cal L} = {\cal L}_{kin}+\sum_{W=X,Y,V,A,u} {\cal L}_{BW} + 
{\cal L}_{vec} + {\cal L}_{0} + {\cal L}_{SB}
\label{genlag}
\end{equation}
In Eq.(\ref{genlag}), ${\cal L}_{kin}$ is the kinetic energy term, 
${\cal L}_{BW}$ 
is the baryon-meson interaction term in which the baryons-spin-0 meson 
interaction term generates the baryon masses. ${\cal L}_{vec}$  describes 
the dynamical mass generation of the vector mesons via couplings to the 
scalar mesons and contain additionally quartic self-interactions of the 
vector fields. ${\cal L}_{0}$ contains the meson-meson interaction terms 
inducing the spontaneous breaking of chiral symmerty as well as a scale 
invariance breaking logarthimic potential. ${\cal L}_{SB}$ describes the 
explicit chiral symmetry breaking. 

To study the hadron properties at finite densities in the present investigation, we use the mean field approximation, where all the meson fields are treated as classical fields. In this approximation, only the scalar and the vector fields contribute to the baryon-meson interaction, ${\cal L}_{BW}$ since for all the other mesons, the expectation values are zero.

The interactions of the scalar mesons and vector mesons with the baryons are given as
\begin{eqnarray}
 {\cal L} _{Bscal} +  {\cal L} _{Bvec} & = & - \sum_{i} \bar{\psi}_{i} \left[ g_{\omega i}  \gamma_{0} \omega + g_{\phi i} \gamma_{0} \phi + g_{\rho i} \gamma_{0} \rho + m_{i}^{*} \right] \psi_{i}
 \label{baryscavec}
\end{eqnarray}
The interaction of the vector mesons, of the scalar fields and the interaction corresponding to the explicitly symmetry breaking in the mean field approximation are given as

\begin{eqnarray}
 {\cal L} _{vec} & = & \frac{1}{2} \left( m_{\omega}^{2} \omega^{2} + m_{\rho}^{2} \rho^{2} + m_{\phi}^{2} \phi^{2} \right) \frac{\chi^{2}}{\chi_{0}^{2}} +  g_{4} \left( \omega^{4} + 6\omega^{2}\rho^{2} + \rho^{4} + 2\phi^{4}\right),
 \label{vec} 
\end{eqnarray}

\begin{eqnarray}
 {\cal L} _{0} & = & -\frac{1}{2} k_{0}\chi^{2} \left( \sigma^{2} + \zeta^{2} + \delta^{2} \right) + k_{1} \left( \sigma^{2} + \zeta^{2} + \delta^{2} \right)^{2} \nonumber\\
&+& k_{2} \left( \frac{\sigma^{4}}{2} + \frac{\delta^{4}}{2} + 3 \sigma^{2} \delta^{2} + \zeta^{4} \right) 
+ k_{3}\chi\left( \sigma^{2} - \delta^{2} \right)\zeta \nonumber\\
&-& k_{4} \chi^{4} - \frac{1}{4} \chi^{4} ln \frac{\chi^{4}}{\chi_{0}^{4}} + \frac{d}{3} \chi^{4} ln \left( \frac{\left( \sigma^{2} - \delta^{2}\right)\zeta }{\sigma_{0}^{2} \zeta_{0}} \left( \frac{\chi}{\chi_{0}}\right) ^{3}\right) ,
\label{lagscal}
\end{eqnarray}

and

\begin{eqnarray}
 {\cal L} _{SB} & = & - \left( \frac{\chi}{\chi_{0}}\right) ^{2} \left[ m_{\pi}^{2} f_{\pi} \sigma + \left(\sqrt{2} m_{k}^{2}f_{k} - \frac{1}{\sqrt{2}} m_{\pi}^{2} f_{\pi} \right) \zeta \right] 
 \label{symmbrk}
\end{eqnarray}
The effective mass of the baryon of species $i$ is given as
\begin{equation}
m^{*}_{i} = - g_{\sigma i}\sigma - g_{\zeta i}\zeta - g_{\delta i} \delta
\label{barymass}
\end{equation}

The baryon-scalar meson interactions generate the baryon masses through 
the coupling of baryons to the nonstrange $\sigma$, strange scalar mesons 
$\zeta$ and also to scalar isovector meson $\delta$. In analogy to the 
baryon-scalar meson coupling there exist two independent baryon-vector 
meson interaction terms corresponding to the F-type (antisymmetric) 
and D-type (symmetric) couplings. Here antisymmetric coupling is used 
because the universality principle \cite{saku69} and vector meson dominance 
model suggest small symmetric coupling. Additionally,  we choose the 
parameters \cite{paper3,isoamss} so as to decouple the strange vector 
field $\phi_{\mu}\sim\bar{s}\gamma_{\mu}s$ from the nucleon, corresponding 
to an ideal mixing between $\omega$ and $\phi$ mesons. A small deviation 
of the mixing angle from ideal mixing \cite{dumbrajs,rijken,hohler1} 
has not been taken into account in the present investigation.

The concept of broken scale invariance leading to the trace anomaly 
in (massless) QCD, $\theta_{\mu}^{\mu} = \frac{\beta_{QCD}}{2g} 
G_{\mu\nu}^{a} G^{a\mu\nu}$, where $G_{\mu\nu}^{a} $ is the gluon 
field strength tensor of QCD, can be mimicked in an effective Lagrangian 
at tree level \cite{sche1} through the introduction of the scale breaking 
potential (last two terms of Eq. (\ref{lagscal}))
\begin{equation}
{\cal L}_{scalebreaking} =  - \frac{1}{4} \chi^{4} ln \frac{\chi^{4}}{\chi_{0}^{4}} + \frac{d}{3} \chi^{4} ln \left( \frac{\left( \sigma^{2} - \delta^{2}\right)\zeta }{\sigma_{0}^{2} \zeta_{0}} \left( \frac{\chi}{\chi_{0}}\right) ^{3}\right) 
\label{scalebreak}
\end{equation}
The effect of these logarithmic terms $\sim \chi^{4} ln \chi$ is to break 
the scale invariance, which leads to the trace of the energy momentum 
tensor as \cite{heide1}

\begin{equation}
\theta_{\mu}^{\mu} = - 4{\cal L} + \chi \frac{\partial {\cal L}}
{\partial \chi}  = -(1 - d)\chi^{4}
\label{tensor1}
\end{equation}
Hence the scalar gluon condensate of QCD $\left( \left\langle \frac{\alpha_{s}}{\pi}G_{\mu\nu}^{a}G^{a\mu\nu}\right\rangle \right)$ is simulated by a scalar dilaton field in the present hadronic model. In the present investigation we use the frozen glue ball limit according to which the scalar dilaton field, $\chi$ has very little dependence on the density of the medium and therefore its expectation value is taken to be constant (equal to the vacuum value, $\chi_{0}$) \cite{paper3}.
The comparison of the trace of the energy momentum tensor arising
from the trace anomaly of QCD with that of the present chiral model
gives the relation of the dilaton field to the scalar gluon condensate.
We have, in the limit of massless quarks \cite{cohen},
\begin{equation}
\theta_{\mu}^{\mu} = \langle \frac{\beta_{QCD}}{2g} 
G_{\mu\nu}^{a} G^{a\mu\nu} \rangle  \equiv  -(1 - d)\chi^{4} 
\label{tensor2}
\end{equation}
The parameter $d$ originates from the second logarithmic term of equation 
(\ref{scalebreak}). To get an insight into the value of the parameter 
$d$, we recall that the QCD $\beta$ function at one loop level, for 
$N_{c}$ colors and $N_{f}$ flavors is given by
\begin{equation}
\beta_{\rm {QCD}} \left( g \right) = -\frac{11 N_{c} g^{3}}{48 \pi^{2}} 
\left( 1 - \frac{2 N_{f}}{11 N_{c}} \right)  +  O(g^{5})
\label{beta}
\end{equation}
In the above equation, the first term in the parentheses arises from 
the (antiscreening) self-interaction of the gluons and the second term, 
proportional to $N_{f}$, arises from the (screening) contribution of 
quark pairs. Equations (\ref{tensor2}) and (\ref{beta}) suggest the 
value of $d$ to be 6/33 for three flavors and three colors, and 
for the case of three colors and two flavors, the value of $d$ 
turns out to be 4/33, to be consistent with the one loop estimate 
of QCD $\beta$ function. These values give the order of magnitude 
about which the parameter $d$ can be taken \cite{heide1}, since one 
cannot rely on the one-loop estimate for $\beta_{\rm {QCD}}(g)$. 
This parameter, along with the other parameters
corresponding to the  scalar Lagrangian density, ${\cal L}_0$ 
given by equation (\ref{lagscal}), are fitted so as to ensure 
extrema in the vacuum for the $\sigma$ and $\zeta$ field 
equations, to  reproduce the vacuum masses of the $\eta$ and $\eta '$ 
mesons, the mass of the $\sigma$ meson around 500 MeV, and pressure, 
p($\rho_0$)=0,
with $\rho_0$ as the nuclear matter saturation density \cite{paper3,amarind}.

The coupled equations of motion for the non-strange scalar field $\sigma$, 
strange scalar field $\zeta$ and the scalar-isovector field $ \delta$  
are derived from the Lagrangian densities given by equations 
(\ref{baryscavec}) to (\ref{symmbrk}) and are given as 
\begin{eqnarray}
k_{0}\chi^{2}\sigma-4k_{1}\left( \sigma^{2}+\zeta^{2}+\delta^{2}\right)\sigma-2k_{2}\left( \sigma^{3}+3\sigma\delta^{2}\right)-2k_{3}\chi\sigma\zeta \nonumber\\
-\frac{d}{3} \chi^{4}\frac{2\sigma}{\sigma^{2}-\delta^{2}}+\left( \frac{\chi}{\chi_{0}}\right) ^{2}m_{\pi}^{2}f_{\pi}-\sum g_{\sigma i}\rho_{i}^{s} = 0 
\label{sigma}
\end{eqnarray}

\begin{eqnarray}
k_{0}\chi^{2}\zeta-4k_{1}\left( \sigma^{2}+\zeta^{2}+\delta^{2}\right)\zeta-4k_{2}\zeta^{3}-k_{3}\chi\left( \sigma^{2}-\delta^{2}\right)\nonumber\\
- \frac{d}{3}\frac{\chi^{4}}{\zeta}+\left(\frac{\chi}{\chi_{0}} \right) ^{2}\left[ \sqrt{2}m_{k}^{2}f_{k}-\frac{1}{\sqrt{2}} m_{\pi}^{2}f_{\pi}\right] -\sum g_{\zeta i}\rho_{i}^{s} = 0 
\label{zeta}
\end{eqnarray}

\begin{eqnarray}
k_{0}\chi^{2}\delta-4k_{1}\left( \sigma^{2}+\zeta^{2}+\delta^{2}\right)\delta-2k_{2}\left( \delta^{3}+3\sigma^{2}\delta\right) +k_{3}\chi\delta \zeta \nonumber\\
+ 2 \frac{d}{3} \chi^{4}\left( \frac{\delta}{\sigma^{2}-\delta^{2}}\right)-\sum g_{\delta i} \rho_{i}^{s} = 0
\label{delta}
\end{eqnarray}
 
In the above, $\rho^{s}_{i}$ are the scalar densities for the baryons, which at zero temperature is given as
\begin{equation}
\rho^{s}_{i} = \gamma_{i} \int_{0}^{k_{Fi}} \frac{d^{3}k}{(2\pi)^{3}} \frac{m_{i}^{*}}{\left( k^{2} + m_{i}^{*2}\right) ^{1/2}} 
\label{scalardens}
\end{equation}
where, $k_{Fi}$ is the fermi momentum of the baryon of species $i$ 
($i = p, n, \Lambda, \Sigma^{\pm,0}, \Xi^{-,0}$) and  $\gamma_{i} = 2$ 
is the spin degeneracy factor \cite{isoamss}.

The above coupled equations of motion are solved to obtain the density 
dependent values of the scalar fields $(\sigma, \zeta$ and $\delta)$, 
in the isospin asymmetric strange hadronic medium at zero temperature. 
The isospin asymmetry of the medium is defined through the 
parameter, $\eta = \left( \rho_{n} - \rho_{p}\right) /\left( 2\rho_{B}\right)$, 
where $\rho_{n}$ and $\rho_{p}$ are the number densities of 
the neutron and proton and $\rho_{B}$ is the baryon density. 
The strangeness fraction, $f_{s}$ of the medium is defined by 
$\frac{\sum_{i} \vert s_{i} \vert \rho_{i}}{\rho_{B}}$, 
where $s_{i}$ is the number of strange quarks of baryon $i$.
In this strange hadronic medium we have both nucleons and hyperons.
The nucleons modified in the strange hadronic matter further interact 
with the $D$ and $\bar{D}$ mesons and modify their properties.

\section{$D$ and $\bar D$ mesons in isospin asymmetric strange hadronic matter}
In this section we study the $D$ and $\bar{D}$ mesons properties in isospin
asymmetric strange hadronic matter. As mentioned earlier, the medium 
modifications of the $D$ and $\bar{D}$ mesons arise due to their 
interactions  with the nucleons, non strange scalar meson $\sigma$ 
and the scalar isovector meson $\delta$. The non strange scalar meson 
$\sigma$ and the scalar isovector meson $\delta$ are modified in the 
medium consisting of nucleons and hyperons. The nucleons modified 
in the strange hadronic medium interact with the $D$ and $\bar{D}$ 
mesons and modify their properties. The interaction Lagrangian density 
is given as
\begin{eqnarray}
\cal L _{DN} & = & -\frac {i}{8 f_D^2} \Big [3\Big (\bar p \gamma^\mu p
+\bar n \gamma ^\mu n \Big) 
\Big({D^0} (\partial_\mu \bar D^0) - (\partial_\mu {{D^0}}) {\bar D}^0 \Big )
+\Big(D^+ (\partial_\mu D^-) - (\partial_\mu {D^+})  D^- \Big )
\nonumber \\
& +&
\Big (\bar p \gamma^\mu p -\bar n \gamma ^\mu n \Big) 
\Big({D^0} (\partial_\mu \bar D^0) - (\partial_\mu {{D^0}}) {\bar D}^0 \Big )
- \Big( D^+ (\partial_\mu D^-) - (\partial_\mu {D^+})  D^- \Big )
\Big ]
\nonumber \\
 &+ & \frac{m_D^2}{2f_D} \Big [ 
(\sigma +\sqrt 2 \zeta_c)\big (\bar D^0 { D^0}+(D^- D^+) \big )
 +\delta \big (\bar D^0 { D^0})-(D^- D^+) \big )
\Big ] \nonumber \\
& - & \frac {1}{f_D}\Big [ 
(\sigma +\sqrt 2 \zeta_c )
\Big ((\partial _\mu {{\bar D}^0})(\partial ^\mu {D^0})
+(\partial _\mu {D^-})(\partial ^\mu {D^+}) \Big )
\nonumber \\
 & + & \delta
\Big ((\partial _\mu {{\bar D}^0})(\partial ^\mu {D^0})
-(\partial _\mu {D^-})(\partial ^\mu {D^+}) \Big )
\Big ]
\nonumber \\
&+ & \frac {d_1}{2 f_D^2}(\bar p p +\bar n n 
 )\big ( (\partial _\mu {D^-})(\partial ^\mu {D^+})
+(\partial _\mu {{\bar D}^0})(\partial ^\mu {D^0})
\big )
\nonumber \\
&+& \frac {d_2}{4 f_D^2} \Big [
(\bar p p+\bar n n))\big ( 
(\partial_\mu {\bar D}^0)(\partial^\mu {D^0})
+ (\partial_\mu D^-)(\partial^\mu D^+) \big )\nonumber \\
 &+&  (\bar p p -\bar n n) \big ( 
(\partial_\mu {\bar D}^0)(\partial^\mu {D^0})\big )
- (\partial_\mu D^-)(\partial^\mu D^+) ) 
\Big ]
\label{ldn}
\end{eqnarray}

In Eq. (\ref{ldn}), the first term is the vectorial Weinberg Tomozawa 
interaction term, obtained from the kinetic term of Eq. (\ref{genlag}). 
The second term is obtained from 
the explicit symmetry breaking term and leads to the attractive interactions, as a function of density, 
for both the $D$ and $\bar{D}$ mesons in the medium. The next three terms of 
above Lagrangian density ($\sim (\partial_\mu {\bar D})(\partial ^\mu D)$)
are known as the range terms. The first range term (with coefficient 
$\big (-\frac{1}{f_D}\big)$) is obtained from the kinetic energy term 
of the pseudoscalar mesons. The second and third range terms $d_{1}$ 
and $d_{2}$ are written for the $DN$ interactions in analogy with 
those written for $KN$ interactions in \cite{isoamss2}. It might be 
noted here that the interaction of the pseudoscalar mesons with the 
vector mesons, in addition to the pseudoscalar meson-nucleon vectorial 
interaction, leads to a double counting in the linear realization of 
chiral effective theories. Further, in the non-linear realization,
such an interaction does not arise in the leading or subleading order, 
but only as a higher order contribution \cite{borasoy}. Hence the 
vector meson-pseudoscalar interactions will not be taken into account
in the present investigation.

The  dispersion relations for the $D$ and $\bar{D}$ mesons are obtained 
by the Fourier transformations of equations of motion. These are given as 
\begin{equation}
-\omega^{2}+\vec{k}^{2}+m_{D}^{2}-\Pi\left(\omega,\vert\vec{k}\vert, 
\rho\right) 
= 0
\label{dispersion}
\end{equation}
where, $m_D$ is the vacuum mass of the $D(\bar D)$ meson and
$\Pi\left(\omega,\vert\vec{k}\vert, \rho\right)$ denotes the self-energy 
of the $D\left( \bar{D} \right) $ mesons in the medium.
The self-energy $\Pi\left( \omega , \vert\vec{k}\vert, \rho\right) $ 
for the $D$ 
meson doublet $ \left( D^{0} , D^{+}\right) $ arising from the interaction 
of Eq.(\ref{ldn}) is given as
\begin{eqnarray}
\Pi (\omega, |\vec k|,\rho) &= & \frac {1}{4 f_D^2}\Big [3 (\rho_p +\rho_n)
\pm (\rho_p -\rho_n) 
\Big ] \omega \nonumber \\
&+&\frac {m_D^2}{2 f_D} (\sigma ' +\sqrt 2 {\zeta_c} ' \pm \delta ')
\nonumber \\ & +& \Big [- \frac {1}{f_D}
(\sigma ' +\sqrt 2 {\zeta_c} ' \pm \delta ')
+\frac {d_1}{2 f_D ^2} (\rho_s ^p +\rho_s ^n)\nonumber \\
&+&\frac {d_2}{4 f_D ^2} \Big (({\rho^s} _p +{\rho^s} _n)
\pm   ({\rho^s} _p -{\rho^s} _n) \Big ) \Big ]
(\omega ^2 - {\vec k}^2),
\label{selfd}
\end{eqnarray}
where the $\pm$ signs refer to the $D^{0}$ and $D^{+}$ mesons, 
respectively, and $\sigma^{\prime}\left( = \sigma - \sigma_{0}\right) $, 
$\zeta_{c}^{\prime}\left( = \zeta_{c} - \zeta_{c0}\right)$, and 
$\delta^{\prime}\left(  = \delta -\delta_{0}\right) $ are the 
fluctuations of the scalar-isoscalar fields $\sigma$, $\zeta_{c}$ and
the scalar-isoscalar field $\delta$ from their 
vacuum expectation values in the strange hyperonic medium. The vacuum expectation value of $\delta$ 
is zero $\left(\delta_{0}=0 \right)$, since a nonzero value for it 
will break the isospin-symmetry of the vacuum. (We neglect here the 
small isospin breaking effect arising  from the mass and charge 
difference of the up and down quarks.) 
We might note here that the interaction of the scalar charm quark condensate 
$\zeta_{c}$ (being made up of heavy charm quarks and charm antiquarks)
leads to very small modifications of the masses \cite{roeder}. So we will 
not consider the medium fluctuations of $\zeta_{c}$. 
In Eq.(\ref{selfd}),
$\rho_{i}$ and $\rho_{i}^{s}$ with $i = p, n$ are the number density and 
the scalar density of the baryon of type $i$. The scalar density, 
$\rho_{i}^{s}$, at zero temperature is already defined in Eq. 
(\ref{scalardens}), whereas the number density, $\rho_{i}$ for the
$i$-th baryon ($i=p,n,\Lambda,\Sigma^{\pm,0},\Xi^{-,0}$), is defined as 
\begin{equation}
\rho_{i} = \gamma_{i} \int_{0}^{k_{Fi}} \frac{d^{3}k}{(2\pi)^{3}}, 
\label{nmdens}
\end{equation}
where $\gamma_i$=2 is the spin degeneracy factor.
  
Similarly, for the  $\bar{D}$ meson doublet 
$\left(\bar{D}^{0},D^{-}\right)$, the self-energy is calculated as 
\begin{eqnarray}
\Pi (\omega, |\vec k|,\rho) &= & -\frac {1}{4 f_D^2}\Big [3 (\rho_p +\rho_n)
\pm (\rho_p -\rho_n) \Big ] \omega\nonumber \\
&+&\frac {m_D^2}{2 f_D} (\sigma ' +\sqrt 2 {\zeta_c} ' \pm \delta ')
\nonumber \\ & +& \Big [- \frac {1}{f_D}
(\sigma ' +\sqrt 2 {\zeta_c} ' \pm \delta ')
+\frac {d_1}{2 f_D ^2} (\rho_s ^p +\rho_s ^n
)\nonumber \\
&+&\frac {d_2}{4 f_D ^2} \Big (({\rho^s} _p +{\rho^s} _n)
\pm   ({\rho^s} _p -{\rho^s} _n) \Big ]
(\omega ^2 - {\vec k}^2),
\label{selfdbar}
\end{eqnarray}
where the $\pm$ signs refer to the $\bar{D}^{0}$ and $D^{-}$ mesons, 
respectively. The optical potentials of the $D$ and $\bar{D}$ mesons are 
obtained using the expression
\begin{equation}
U(\omega, k) = \omega(k) - \sqrt{k^{2} + m_{D}^{2}}
\end{equation}
where $m_{D}$ is the vacuum mass for the $D(\bar{D})$ meson and 
$\omega(k)$ is the momentum-dependent energy of the $D(\bar{D})$ meson.
\section{Results and Discussions}

In this section we present the results and discussions of our investigation 
of the in-medium properties of $D$ and $\bar{D}$ mesons in isospin asymmetric 
strange hadronic matter. We have generalized the chiral $SU(3)$ model to 
$SU(4)$ to include the interactions of the charmed mesons. The present 
calculations use the following model parameters: $k_{0} = 2.54, 
k_{1} = 1.35, k_{2} = -4.78, k_{3} = -2.77$, $k_{4} = -0.22$ 
and $d =  0.064$, which are the parameters occurring in the scalar meson 
interactions defined in equation (\ref{lagscal}). The vacuum values of the
scalar isoscalar fields, $\sigma$ and $\zeta$ and the dilaton field $\chi$ 
are $-93.3$ MeV, $-106.6$ MeV and $409.8$ MeV respectively. The values, 
$g_{\sigma N} = 10.6$ and $g_{\zeta N} = -0.47$ are determined by fitting 
vacuum baryon masses. The other parameters fitted to the asymmetric 
nuclear matter saturation properties in the mean-field approximation 
are: $g_{\omega N}$ = 13.3, $g_{\rho p}$ = 5.5, $g_{4}$ = 79.7, 
$g_{\delta p}$ = 2.5, $m_{\zeta}$ = 1024.5 MeV, $ m_{\sigma}$ = 466.5 MeV 
and $m_{\delta}$ = 899.5 MeV. The coefficients $d_{1}$ and $d_{2}$, 
calculated from the empirical values of the $KN$ scattering lengths 
for $I = 0$ and $I = 1$ channels, are $2.56/m_{K}$ and $0.73/m_{K}$, 
respectively \cite{isoamss2}.

The $D$ and $\bar{D}$ mesons properties in isospin asymmetric strange 
hyperonic matter are modified due to their interactions with nucleons, 
the scalar meson $\sigma$ and scalar isovector meson $\delta$. As discussed 
earlier the non-strange scalar meson $\sigma$ and scalar isovector meson 
$\delta$ are modified in the strange hadronic medium consisting of nucleons 
and hyperons. The nucleons modified in the strange hadronic matter, interact 
with the $D$ and $\bar{D}$ mesons. In our present investigation, the 
Weinberg Tomozawa term introduces the isospin asymmetry in the medium 
through the isospin asymmetry in proton and neutron number densities. 
The scalar exchange term and first range term (term with coefficient 
($-\frac{1}{f_{D}}$)) also contribute to the isospin asymmetry of the medium 
through the scalar isovector meson $\delta$. The $d_{2}$-term in the 
interaction Lagrangian density given by Eq. (\ref{ldn}) also introduces 
the isospin asymmetry in the $D (D^{0}, D^{+})$ and $\bar{D} 
(\bar{D^{0},D^{-})}$ meson doublets.
  
\begin{figure}
\includegraphics[width=18cm,height=16cm]{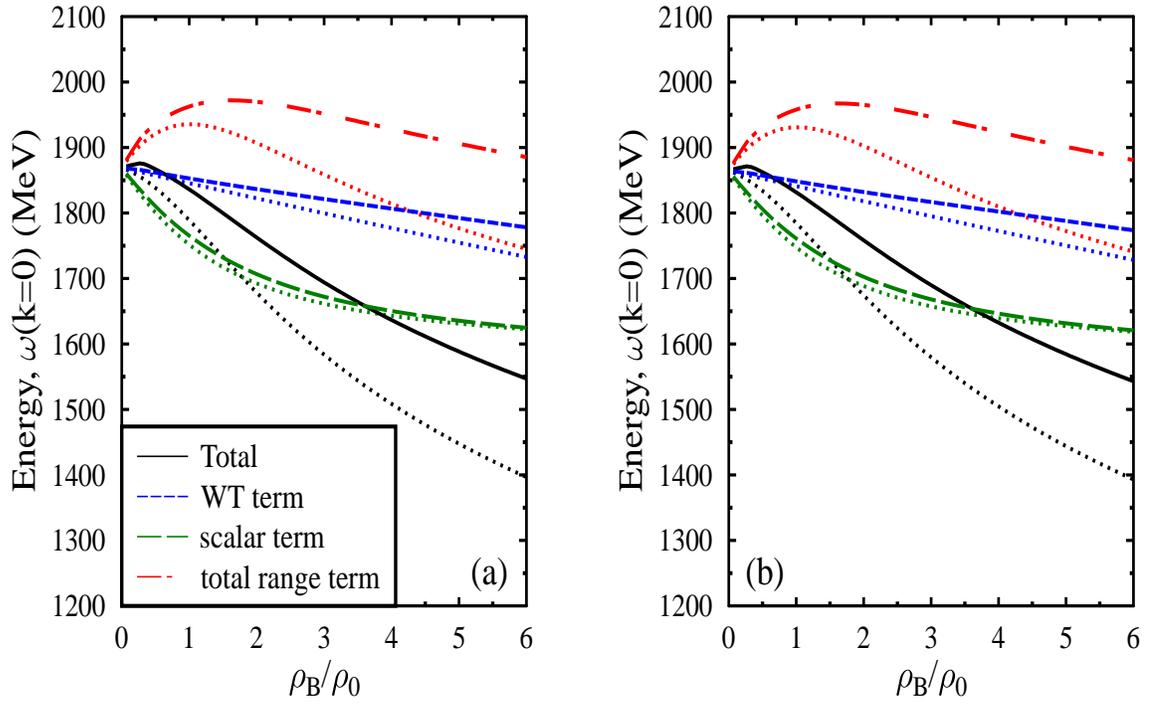} 
\caption{(Color online) The various contributions to $D$ meson energies 
at zero momentum in isospin symmetric strange hadronic medium ($\eta = 0$) 
for (a) $D^{+}$ and for (b) $D^{0}$ in MeV plotted as functions of baryon 
density in units of nuclear matter saturation density, $\rho_{B}/\rho_{0}$, 
shown for the strangeness fraction $f_{s} = 0.5$ and compared with the case 
of $f_{s} =0$ (dotted line).}
\label{fig.1}
\end{figure}

\begin{figure}
\includegraphics[width=18cm,height=16cm]{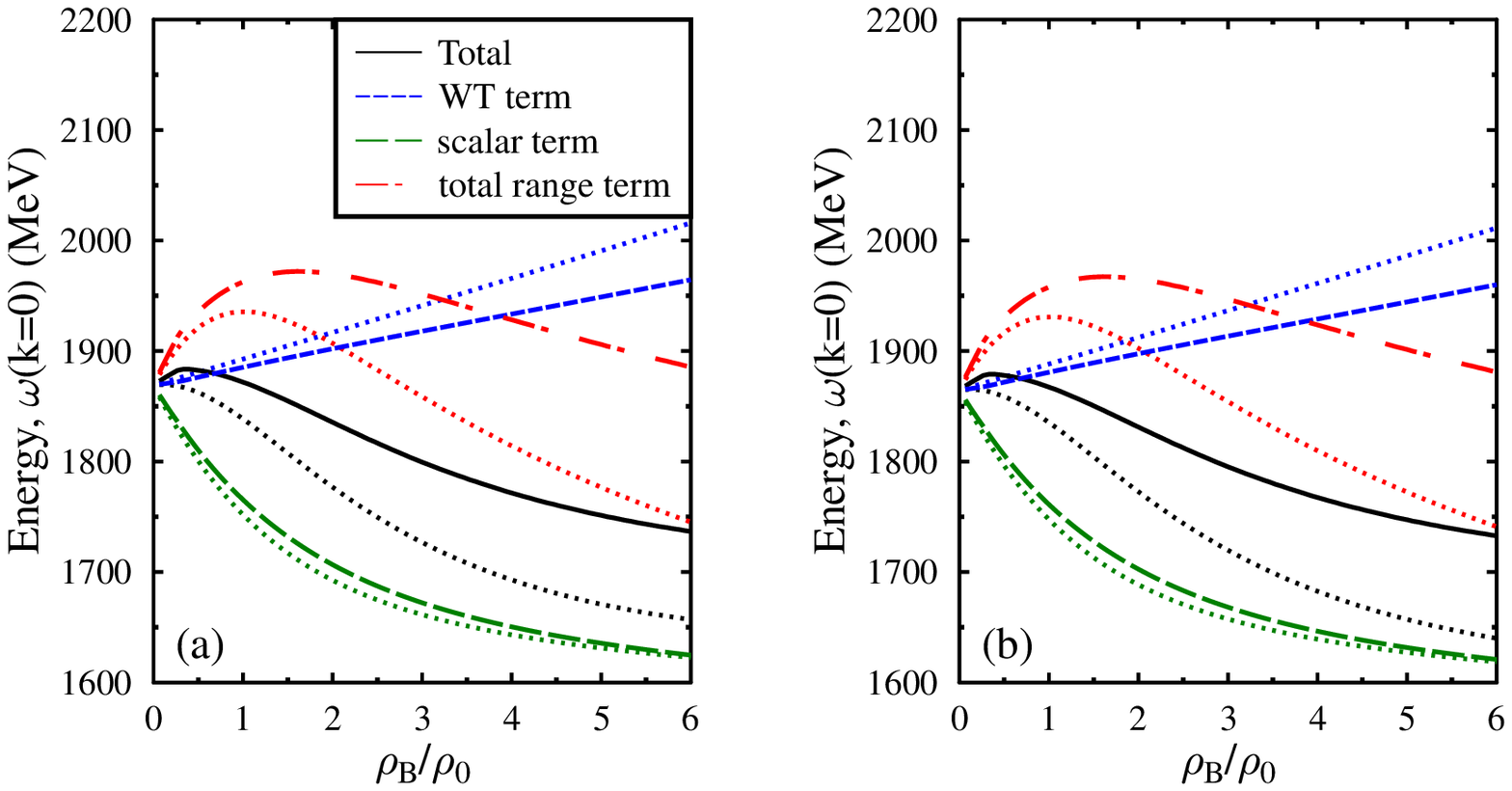} 
\caption{(Color online) The various contributions to $\bar{D}$ meson energies 
at zero momentum in isospin symmetric strange hadronic medium ($\eta = 0$) 
for (a) $D^{-}$ and for (b) $\bar{D^{0}}$ in MeV plotted as functions of 
baryon density in units of nuclear matter saturation density, $\rho_B/\rho_0$, 
shown for the strangeness fraction $f_{s} = 0.5$ and compared with the case 
of $f_{s} =0$ (dotted line).}
\label{fig.2}
\end{figure}

First, we study the effect of strangeness fraction, $f_{s}$ of the medium 
on the energies of $D$ and $\bar{D}$ mesons arising due to the various
terms of the Lagrangian density given by equation (\ref{ldn}).
In figures \ref{fig.1} and \ref{fig.2}, we show the variation of 
the energies of $D$ and 
$\bar{D}$ mesons at zero momentum, with density in the isospin symmetric 
medium ($\eta = 0$). The results are shown for values of the strangeness 
fraction, $f_{s} = 0$ and $0.5$. For a given value of isospin 
asymmetry parameter, $\eta$ and the strangeness fraction, $f_{s}$, 
the masses of $D^{+}$ and $D^{0}$ mesons are seen to decrease with 
increase in the density. For a 
given value of density, $\rho_{B}$ and isospin asymmetry parameter, $\eta$, as
we move from nuclear medium ($f_s$=0) to the hyperonic matter, the 
attractive contribution to the in-medium energies of $D$ mesons 
from the Weinberg-Tomozawa term becomes smaller and the repulsive 
contribution arising from this Weinberg-Tomozawa term
becomes smaller for the $\bar{D}$ mesons. To have an 
understanding of this behavior of $D$ and $\bar{D}$ mesons with the
strangeness fraction, $f_{s}$, let us consider the term corresponding
to the contribution of Weinberg-Tomozawa term to the self energy of 
the $D$ mesons at low densities. The D meson self-energy arising from
the Weinberg-Tomozawa interaction, $\Pi_{WT} (\omega, |\vec{k}|)$, is given
by the first term of Eq. (\ref{selfd}), and, at low densities, this turns
out to be much smaller than ($k^{2} + m_{D}^{2}$). One can then, as
a first approximation, replace $\Pi_{WT} (\omega, |\vec{k}|)$ 
by $\Pi_{WT} (m_{D}, |\vec{k}|)$ and solve for the dispersion relation 
given by Eq. (\ref{dispersion}). Confining our attention to the 
Weinberg Tomozawa interaction only, one finds that the energies 
of $D$ mesons($D^{0}$ and $D^{+}$)  are given as \cite{amarind}
\begin{equation}
\omega(\vert\vec{k}\vert) \simeq \left( \vert\vec{k}\vert ^2 
+ {m_D} ^2 \right) ^{1/2} - \frac{1}{8f_{D}^{2}} 
\frac{m_{D}}{\sqrt{k^{2} + m_{D}^{2}}} \left[ 3\left( \rho_{p} 
+ \rho_{n} \right) \pm \left( \rho_{p} - \rho_{n} \right) \right].
\label{selfwt}
\end{equation}
Due to the presence of hyperons in the strange medium, for a fixed value 
of baryon density, $\rho_{B}$, the values of the proton and neutron
densities, $\rho_{p}$ and $\rho_{n}$ are smaller in the 
hyperonic matter (nonzero $f_{s}$) as compared to the non-strange
($f_{s} = 0$) medium. This causes an increase in the energy of 
the $D$ mesons with the strangeness of the medium, as can be seen from 
Eq. (\ref{selfwt}). For $\bar{D}$ meson the sign of the Weinberg-Tomozawa
term is opposite as compared to that of the $D$ meson, as can be seen 
from Eq. (\ref{selfdbar}), and this leads to a smaller repulsive contribution
for the the energy of $\bar{D}$ mesons arising from the Weinberg-Tomozawa
interaction with strangeness fraction, for a fixed value of density 
and isospin asymmetry of the medium. 


For a given value of isospin asymmetry parameter, $\eta$ and the strangeness 
fraction, $f_{s}$, the scalar meson exchange term 
has attractive contributions to the energies 
of both $D$ and $\bar{D}$ mesons, as can be seen from the second terms
of the equations (\ref{selfd}) and (\ref{selfdbar}). From figures 
\ref{fig.1} and \ref{fig.2}, 
we observe that in isospin symmetric medium, $\eta = 0$, the energies
of $D$ and $\bar{D}$ mesons, due to the scalar exchange term, increase 
as we move from the value of the strangeness fraction, $f_{s} = 0$ to $0.5$.
This can be understood from the following relation of the  energy of $D$ 
and $\bar{D}$ mesons in the symmetric strange hadronic medium at zero 
momentum due to scalar meson exchange term only
 \begin{equation}
\omega(\vert\vec{k}\vert = 0) 
 = m_{D}\left[ 1 - \frac{\sigma^{'}}{2f_{D}}\right] ^{1/2} 
\label{selfsca}
\end{equation}
Note that we have neglected the small fluctuations in the $\zeta_{c}$ field.
As we move from the non-strange medium to strange medium, the magnitude 
of the $\sigma$ field increases and hence, the fluctuation, $\sigma^{'}
(=\sigma -\sigma_0)$ becomes smaller in the strange medium as compared 
to its value in the nuclear medium. This leads to an increase in the 
energy of $D$ and $\bar{D}$ mesons due to scalar meson exchange term 
as we move from symmetric non-strange hadronic medium ($f_{s} = 0$) 
to symmetric strange hadronic medium (nonzero $f_{s}$).

Among the three range terms, the first range 
term (term with coefficient $(-\frac{1}{f_{D}})$) 
has repulsive contributions to the energies of 
$D$ and $\bar{D}$ mesons. However, as we move from the nuclear
medium ($f_s$=0) to hyperonic matter (nonzero $f_s$), the magnitude
of the first range term decreases.
This is due to the fact that there is an increase in the 
magnitude of the $\sigma$ field in moving from 
non-strange hadronic medium to strange hadronic medium, 
which leads to a smaller value of $\sigma '$, the fluctuation of $\sigma$.
The second ($d_{1}$ term) and third ($d_{2}$ term) range terms 
have attractive contributions to the energies of $D$ and $\bar{D}$ mesons. 
However, the attractive contributions are lessened when we go from
nuclear matter ($f_s$=0) to hyperonic matter,
due to smaller values of the proton and neutron scalar densities
$\rho_{s}^{p}$ and $\rho_{s}^{n}$ in the strange 
medium as compared to the non strange medium, for a given 
value of baryon density.
The combined effect of all the range terms is observed to
give an increase in the energies of the $D$ and $\bar D$ mesons
with increase in strangeness of the hadronic medium, as can be seen 
from figures \ref{fig.1} and \ref{fig.2}.

\begin{figure}
\includegraphics[width=16cm,height=16cm]{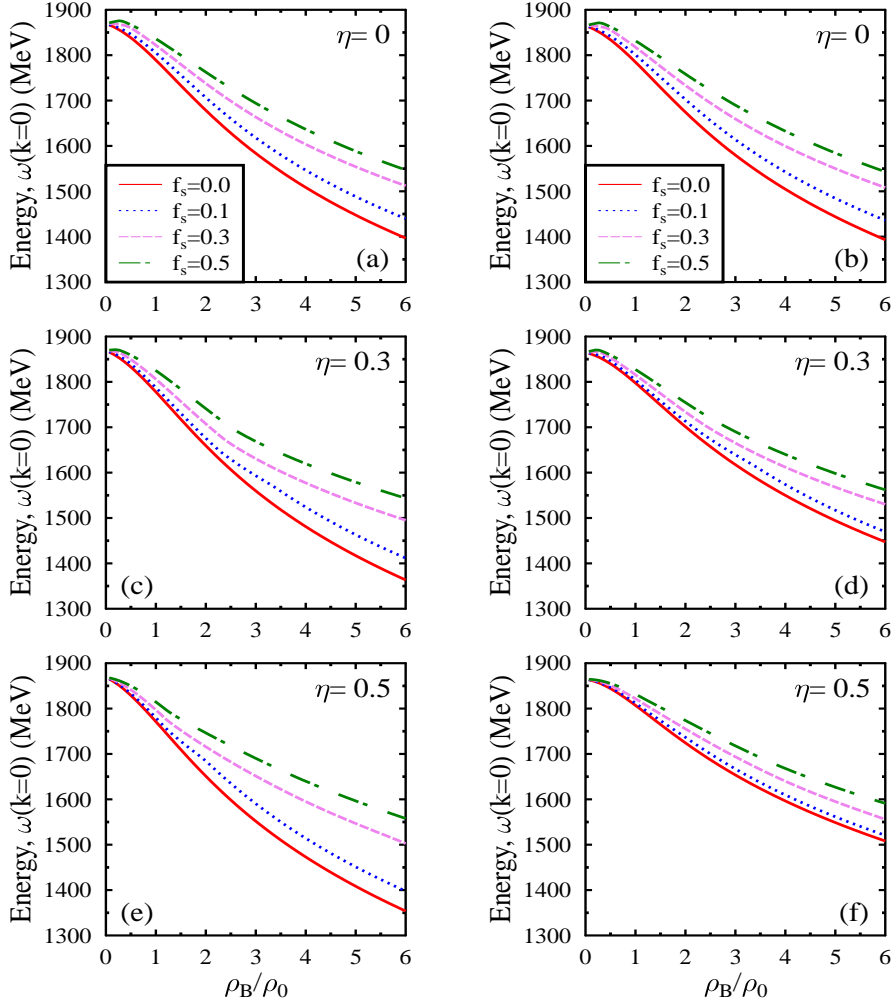} 
\caption{(Color online) The energies of $D^{+}$ mesons ((a),(c) and (e)) 
and of $D^{0}$ mesons ((b),(d) and (f)), at momentum $k = 0$, versus 
the baryon density (in units of nuclear saturation density), 
$\rho_{B}/\rho_{0}$, for different values of the strangeness fractions ($f_{s} = 0, 0.1, 0.3, 0.5$) and for a given value of isospin asymmetry parameter
($\eta = 0, 0.3$ and $0.5$).} 
\label{fig.3}
\end{figure}

In figure \ref{fig.3}, we show the variation of energies of 
D-mesons ($D^{+}, D^{0}$) at zero  momentum, with
density, for different values of isospin asymmetry 
parameters ($\eta$ = 0, 0.3 and 0.5). At each value 
of isospin asymmetry parameter, $\eta$, the results are
shown for strangeness fractions, $f_{s} = 0, 0.1, 0.3$ 
and $0.5$. The $D$ and $\bar{D}$ mesons properties in 
isospin asymmetric nuclear matter at zero 
temperature have been studied in Ref. \cite{amarind}. 
For a given value of density, as we move from isospin 
  symmetric nuclear medium ($\eta = 0$) to isospin asymmetric
   nuclear medium (finite $\eta$), the $D^{+}$ mesons 
   feel a drop in the mass whereas the mass of $D^{0}$ 
   mesons increases. For example, at density 
   $\rho_{B} = \rho_{0}$, $f_{s} = 0$, the mass 
   of $D^{+}$ drops by $18$ MeV whereas the mass of
    $D^{0}$ meson increases by $22$ MeV as we move 
  from $\eta = 0$ to $\eta = 0.5$. At baryon density
 $\rho_{B} = 4\rho_{0}$, $f_{s} = 0$, the energy 
of $D^{+}$ meson decreases by $36$ MeV whereas the
energy of $D^{0}$ mesons increases by $92$ MeV in 
going from $\eta = 0$ to $\eta = 0.5$. At strangeness
fraction $f_{s} = 0.5$, as we move from 
$\eta = 0$ to $\eta = 0.5$, the energy of $D^{+}$ mesons
decreases by $21$ MeV at density $\rho_{B} = \rho_{0}$
whereas at density $4\rho_{0}$ it increases by 3 MeV.
We observe that, at baryon density, $\rho_{B} = \rho_{0}$, as we move from 
isospin symmetric medium ($\eta =0$) to isospin asymmetric medium 
($\eta = 0.5$), there is a larger drop in the energy of $D^{+}$ meson 
at strangeness fraction $f_{s} = 0.5$ as compared to $f_{s} = 0$. 
This is due to the larger drop in the energy of $D^{+}$ meson, 
given by scalar meson exchange term, as a function of isospin asymmetry 
of the medium at strangeness fraction, $f_{s} = 0.5$ as compared to 
$f_{s} =0$. However, the increase in the energy of $D^{+}$ mesons 
as a function of isospin asymmetry of the medium at baryon density, 
$\rho_{B}= 4\rho_{0}$ and strangeness fraction, $f_{s} = 0.5$, 
is because of larger increase in the energy of $D^{+}$ mesons, 
given by the first range term (term with coefficient 
$\frac{-1}{f_{D}}$) at $f_{s} = 0.5$ as compared to $f_{s} = 0$. 
For $D^{0}$ mesons, at strangeness fraction, $f_{s} = 0.5$, 
as we move from $\eta = 0$ to $\eta = 0.5$
the energy increases by $0.14$ MeV and $35$ MeV at 
a baryon density of $\rho_{0}$ and $4\rho_{0}$ respectively.
 
For a given value of density and isospin asymmetry parameter, $\eta$, 
the energies of $D^{+}$ and $D^{0}$ mesons increase with increase 
in the strangeness fraction, $f_{s}$, of the medium. For example, 
at nuclear saturation density $\rho_{0}$, isospin asymmetry parameter 
$\eta = 0$, as we move from $f_{s} = 0$ to $0.5$, the energy of both 
$D^{+}$ and $D^{0}$ mesons increases by $47$ MeV. At $\rho_{B} = 4\rho_{0}$, 
$\eta = 0$ , as we move from $f_{s} = 0$ to $0.5$, the energies of both $D^{+}$
and $D^{0}$ mesons increase by about $128$ MeV. In isospin asymmetric medium 
($\eta = 0.5$), the energy of $D^{+}$ increases by $43$ MeV at $\rho_0$ 
and by $167$ MeV at $4\rho_{0}$ whereas the energy of $D^{0}$ meson 
increases by $25$ and $71$ MeV at $\rho_{0}$ and $4\rho_{0}$ respectively. 
It may be noted that as a function 
of strangeness fraction the increase in the energy of $D^{+}$ mesons 
is larger as compared to $D^{0}$ mesons. 

For a given value of isospin asymmetry and strangeness fraction, the energy 
of $D$ mesons ($D^{0}, D^{+}$) is found to decrease with increase in the 
density of the medium. In isospin symmetric nuclear medium ($\eta = 0$), 
the energy of $D^{+} (D^{0})$ meson at zero momentum decreases by 
80 (79.5) MeV and 360 (359.5) 
MeV at $\rho_{0}$ and $4\rho_{0}$ respectively from its vacuum value.
In isospin symmetric nuclear medium ($\eta = 0$), at the value of the
strangeness fraction $f_{s} = 0.5$, the energy of $D^{+} (D^{0})$ meson 
at $|\vec k|=0$ decreases by $33 (32.5)$ and $232 (231.5)$ MeV 
at $\rho_{0}$ and $4\rho_{0}$ respectively from its vacuum value.
In isospin asymmetric nuclear medium with $\eta = 0.5$, the energy 
of $D^{+} (D^{0})$ meson decreases by 98 (57) MeV and 397 (268) MeV 
at densities of $\rho_{0}$ and $4\rho_{0}$ respectively. 
At the same value of $\eta$, but with $f_{s} = 0.5$, the energy 
of $D^{+} (D^{0})$ is observed to decrease by 54 (33) MeV and 229 (196) MeV 
at densities of $\rho_{0}$ and $4\rho_{0}$ respectively from its vacuum value. 
The smaller drop in the masses of $D$ mesons in the hyperonic medium
is because of an increase in the mass of $D$ mesons with strangeness 
fraction of the medium, as has already been discussed.     
 
\begin{figure}
\includegraphics[width=16cm,height=16cm]{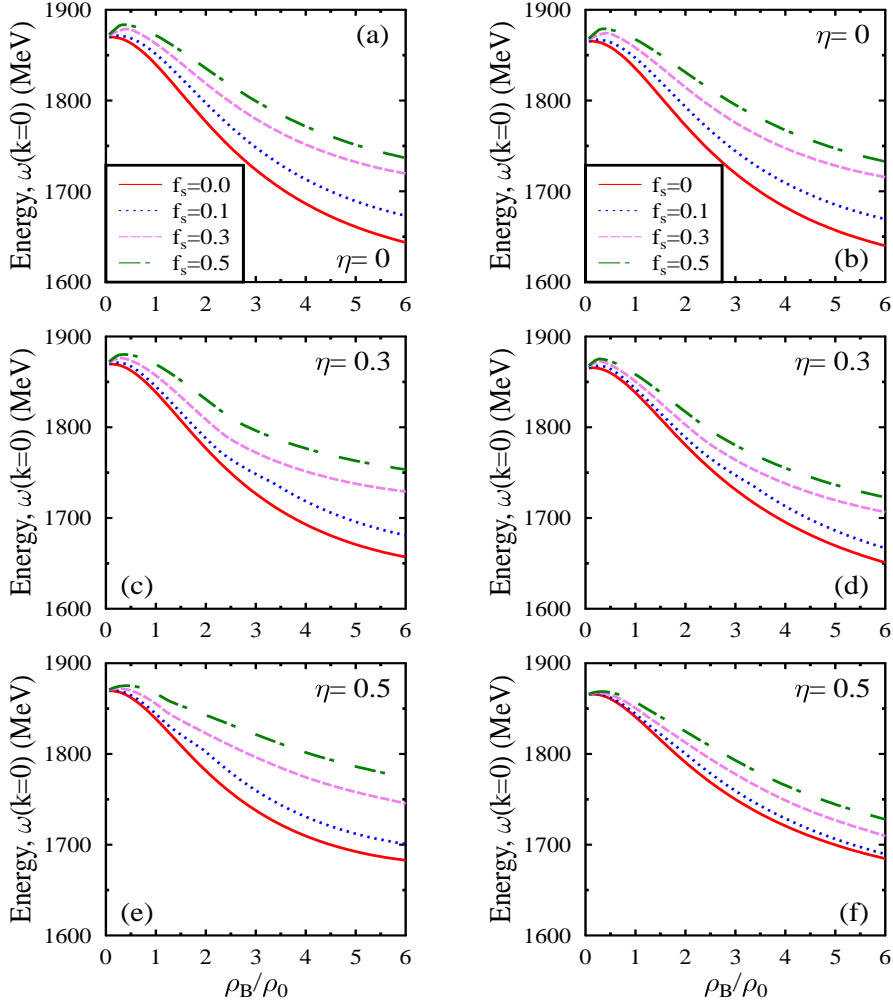} 
\caption{(Color online) The energies of $D^{-}$ mesons ((a),(c) and (e)) 
and of $\bar{D^{0}}$ mesons ((b),(d) and (f)), at momentum $k = 0$, versus 
the baryon density (in units of nuclear saturation density), 
$\rho_{B}/\rho_{0}$, for different values of the strangeness fractions ($f_{s} = 0, 0.1, 0.3, 0.5$) and for a given value of isospin asymmetry parameter 
($\eta = 0, 0.3$ and $0.5$).} 
\label{fig.4}
\end{figure}

In figure \ref{fig.4}, we show the variation of the energies of $\bar{D}$ 
mesons ($D^{-}, \bar{D^{0}}$) at zero  momentum, with density, for different 
values of isospin asymmetry parameter ($\eta = 0, 0.3$ and $0.5$). 
For each value of the isospin asymmetry parameter $\eta$, the results 
are shown for different values of the strangeness fraction, $f_{s}$. 
For a given value of density and isospin asymmetry of the medium, 
similar to the $D$ mesons, the energies of $\bar{D}$ mesons also increase 
with increase in the strangeness fraction $f_{s}$. In isospin symmetric 
medium ($\eta = 0$), for a given value of density, as we move from 
$f_{s} = 0$ to $f_{s} = 0.5$, the energy of $D^{-}$ meson increases 
by $32$ MeV at $\rho_{0}$ and $86$ MeV at $4\rho_{0}$.  At isospin 
asymmetry parameter $\eta = 0.5$, the energy of $D^{-}$ increases 
by $28$ and $62$ MeV at density $\rho_{0}$ and $4\rho_{0}$ respectively. 
In isospin symmetric medium ($\eta = 0$), as we move from $f_{s} = 0$ 
to $0.5$, the energy of $\bar{D^{0}}$ meson increases by $32$ MeV and 
$84$ MeV at $\rho_{0}$ and $4\rho_{0}$ respectively. At $\eta = 0.5$, 
the energy of $\bar{D^{0}}$ meson increases by $17$ MeV and $45$ MeV 
at density $\rho_{0}$ and $4\rho_{0}$ respectively.
The increase in the energy of $\bar{D}$ mesons as a function of the 
strangeness fraction of the medium is because  the range term,
which is repulsive, dominates over the Weinberg-Tomozawa and scalar
exchange terms of the Lagrangian density. 
The $D^{-}$ meson is observed to have  a larger increase in the mass
with increase in the strangeness fraction. This is because the total 
range term has larger repulsive contributions to the 
energy of $D^{-}$ mesons as compared to $\bar{D^{0}}$ mesons as we increase
the strangeness fraction of the hadronic medium.  

The density effects on the in-medium masses of  $\bar{D}$ mesons are
observed to be quite appreciable. In isospin symmetric nuclear medium 
($\eta = 0$, $f_{s} = 0$), the energy of $D^{-} (\bar{D^0})$ meson 
is seen to decrease by $29$ ($30.5$) and $183$ ($181.5$) MeV from 
its vacuum value at densities of $\rho_{0}$ and $4\rho_{0}$ respectively. 
As already mentioned, 
the in-medium energies of $\bar{D}$ mesons increase with increase 
in the strangeness fraction of the medium. At a value of the 
strangeness fraction $f_{s} = 0.5$, we observe a small increase 
in the in-medium energies of $\bar{D}$ mesons with density at small 
baryon densities. For example, in isospin 
symmetric nuclear medium, $\eta = 0$, at strangeness fraction 
$f_{s} = 0.5$, the energy of $D^{-}$ and $\bar{D^0}$ mesons increase 
upto a density of about $0.4\rho_{0}$, above which they start
decreasing with further increase in the density of the medium. 
This behaviour of $\bar{D}$ mesons is because of the fact that the
scalar meson exchange term and the attractive range terms ($d_{1}$ and 
$d_{2}$ terms) become less attractive with increase in the strangeness 
fraction of the medium. Because of this, at small densities the 
in-medium masses of the $\bar{D}$ mesons become larger than their 
vacuum values, since the repulsive vectorial interaction dominates
over the attractive scalar meson and attractive $d_1$ and $d_2$
terms. However,  with further increase in the baryon density the 
the masses of the $\bar D$ mesons decrease with density
although the decrease is less for non-zero values of the strangeness
fraction, $f_s$, as compared to the nuclear matter case ($f_s$=0), 
as can be seen from figure \ref{fig.4}.
In isospin symmetric strange hadronic medium, with $f_s$=0.5, 
at the nuclear saturation density, $\rho_{B} = \rho_{0}$, 
the in-medium energy of the $D^{-}$ and $\bar{D^{0}}$ mesons 
are observed to increase by $3$ MeV and $2.5$ MeV respectively 
from their vacuum values, whereas these masses are seen to drop
by about 30.4 MeV and 34.4 MeV, for the nuclear medium ($f_s$=0). 
At baryon density $\rho_{B} = 4\rho_{0}$, for isospin symmetric
hyperonic matter with $f_s$=0.5, the energy of the $D^{-} 
(\bar{D^{0}})$ meson is observed to drop by $97.5 (97)$ MeV,
whereas the mass drop is 183 (182) MeV for $f_s$=0.

In the present investigation, we observe that the effect of strangeness 
fraction of the medium is to increase the energies of the $D$ and $\bar{D}$ 
mesons and the effect is seen to be more appreciable for 
the $D$ mesons as compared to the $\bar{D}$ mesons. 
The reason for this behaviour can be understood in the following way.
The magnitude of the Weinberg-Tomozawa term, which is attractive 
(repulsive) for the in-medium masses of the $D(\bar D)$ mesons is 
observed to be lessened when we increase the strangeness fraction 
in the hadronic medium, due to the presence of hyperons
in the medium, which gives smaller values for $\rho_p$ and $\rho_n$,
for a given baryon density, and hence the magnitude of the
contribution to the Weinberg-Tomozawa term is smaller. 
On the other hand, in the magnitude of the 
range terms, there is seen to be an increase for both the D and $\bar D$ 
mesons, as can be observed from figures \ref{fig.1} and \ref{fig.2}.
This is due to smaller values of $\rho_s^p$ and $\rho_s^n$ in the presence
of hyperons, for a given density, leading to smaller attractive contributions
from the $d_1$ and $d_2$ terms of the range term. The scalar exchange 
term is seen to decrease in magnitude for both $D$ and $\bar D$
mesons, though the change is observed to be small. The difference 
in the effect of increasing strangeness in the medium on the masses of
the $D (\bar D)$ mesons is due to the positive (negative) contribution
arising from the Weinberg-Tomozawa term, leading to a 
larger increase of the $D$ meson masses as compared to the 
$\bar D$ meson masses, with strangeness of the medium.

\begin{figure}
\includegraphics[width=16cm,height=16cm]{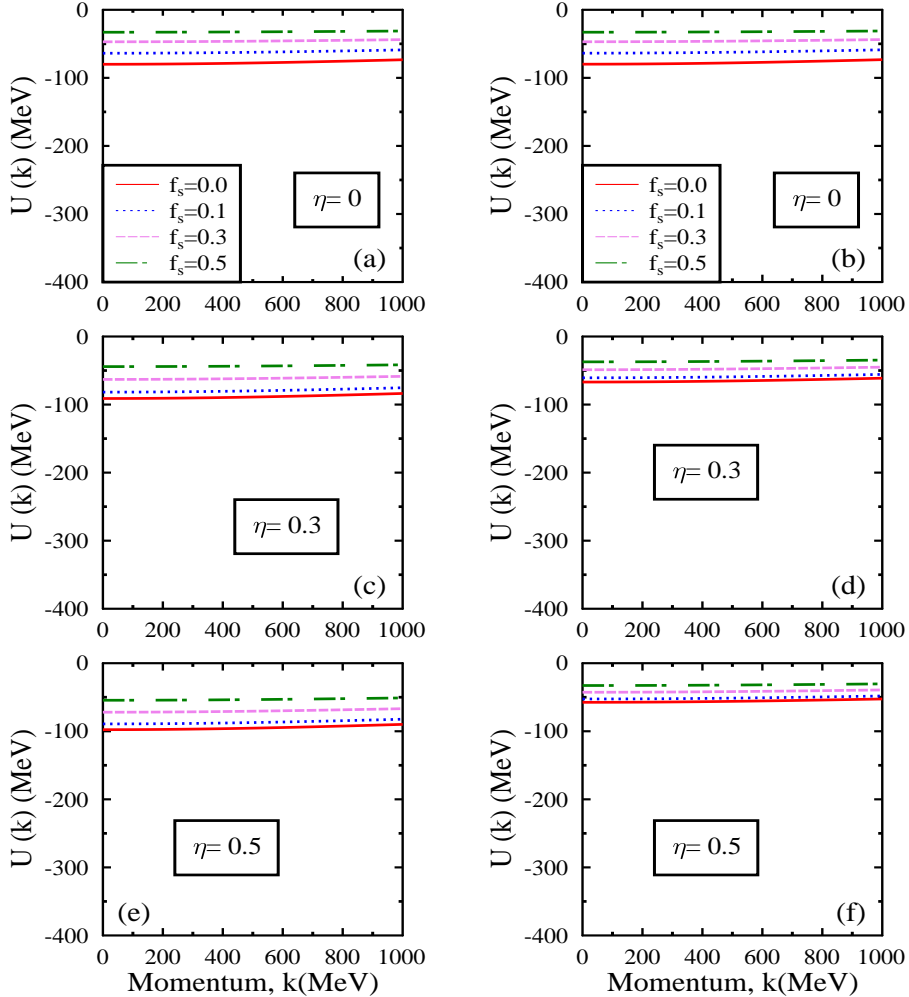} 
\caption{(Color online) The optical potential of $D^{+}$ meson 
(a,c and e) and of $D^{0}$ meson (b,d and f), 
are plotted as functions of momentum for $\rho_{B}=\rho_0$, 
for different values of the strangeness fractions 
($f_{s} = 0, 0.1, 0.3, 0.5$) and for a given value 
of isospin asymmetry parameter ($\eta = 0, 0.3$ and $0.5$).} 
\label{fig.5}
\end{figure}

\begin{figure}
\includegraphics[width=16cm,height=16cm]{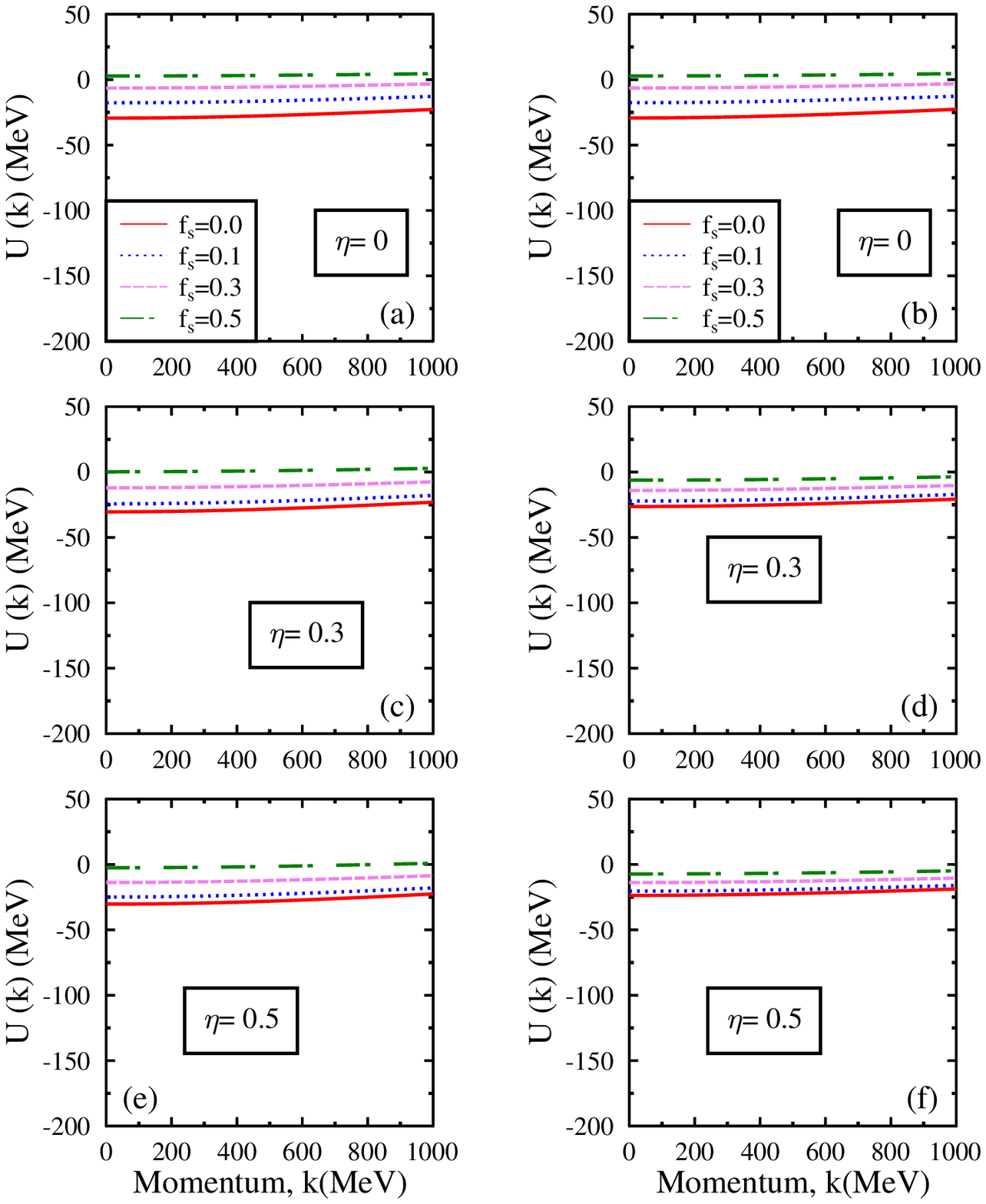} 
\caption{(Color online) The optical potential of $D^{-}$ meson 
(a,c and e) and of $\bar{D^{0}}$ meson (b,d and f), 
are plotted as functions of momentum for $\rho_{B}=\rho_0$, 
for different values of the strangeness fractions 
($f_{s} = 0, 0.1, 0.3, 0.5$) and for a given value  
of isospin asymmetry parameter  ($\eta = 0, 0.3$ and $0.5$).} 
\label{fig.6}
\end{figure}

\begin{figure}
\includegraphics[width=16cm,height=16cm]{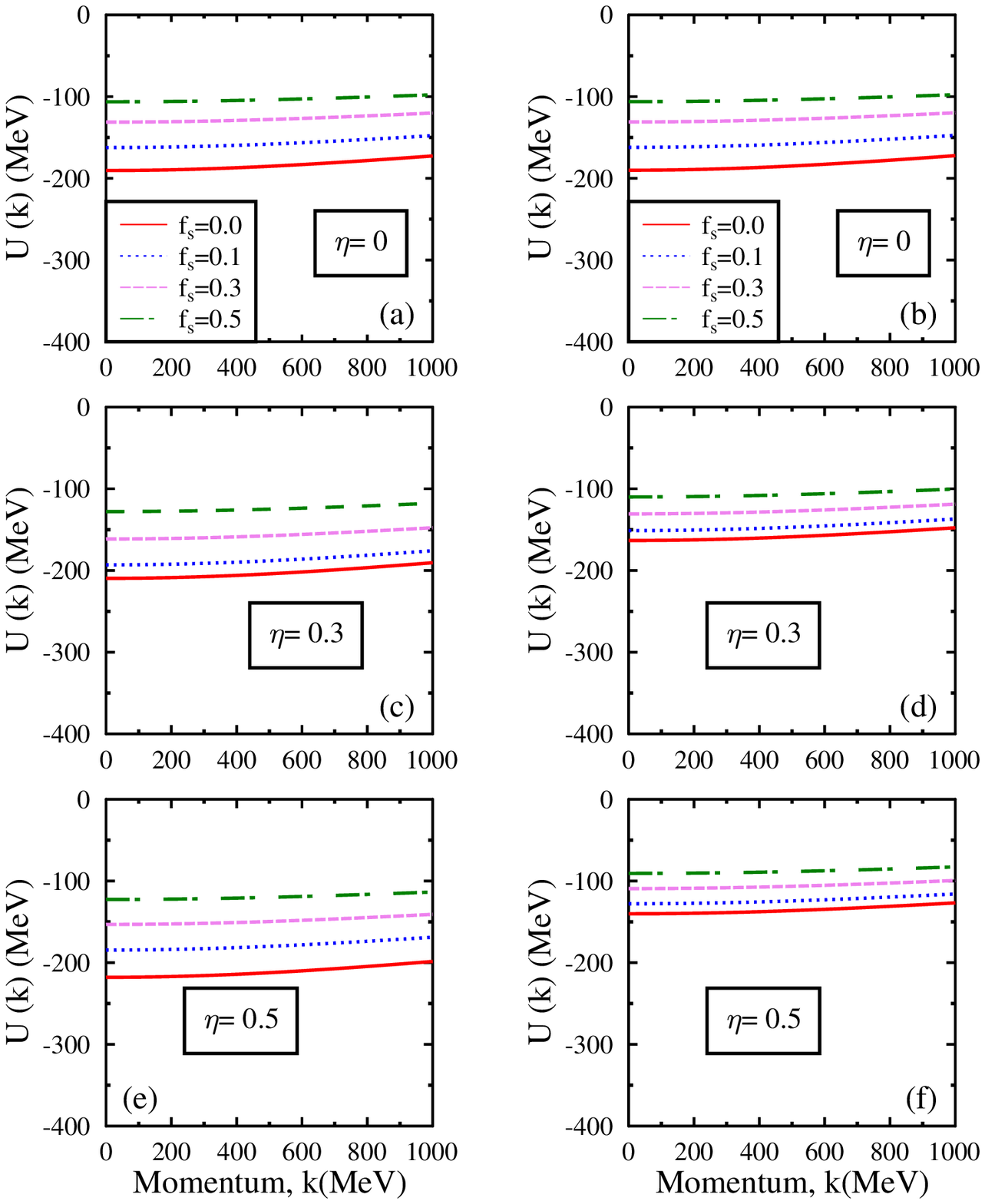} 
\caption{(Color online) The optical potential of $D^{+}$ meson 
(a,c and e) and of $D^{0}$ meson (b,d and f), 
are plotted as functions of momentum for $\rho_{B} = 2\rho_0$, 
for different values of the strangeness fractions 
($f_{s} = 0, 0.1, 0.3, 0.5$) and for a given value 
of isospin asymmetry parameter ($\eta = 0, 0.3$ and $0.5$).} 
\label{fig.7}
\end{figure}

\begin{figure}
\includegraphics[width=16cm,height=16cm]{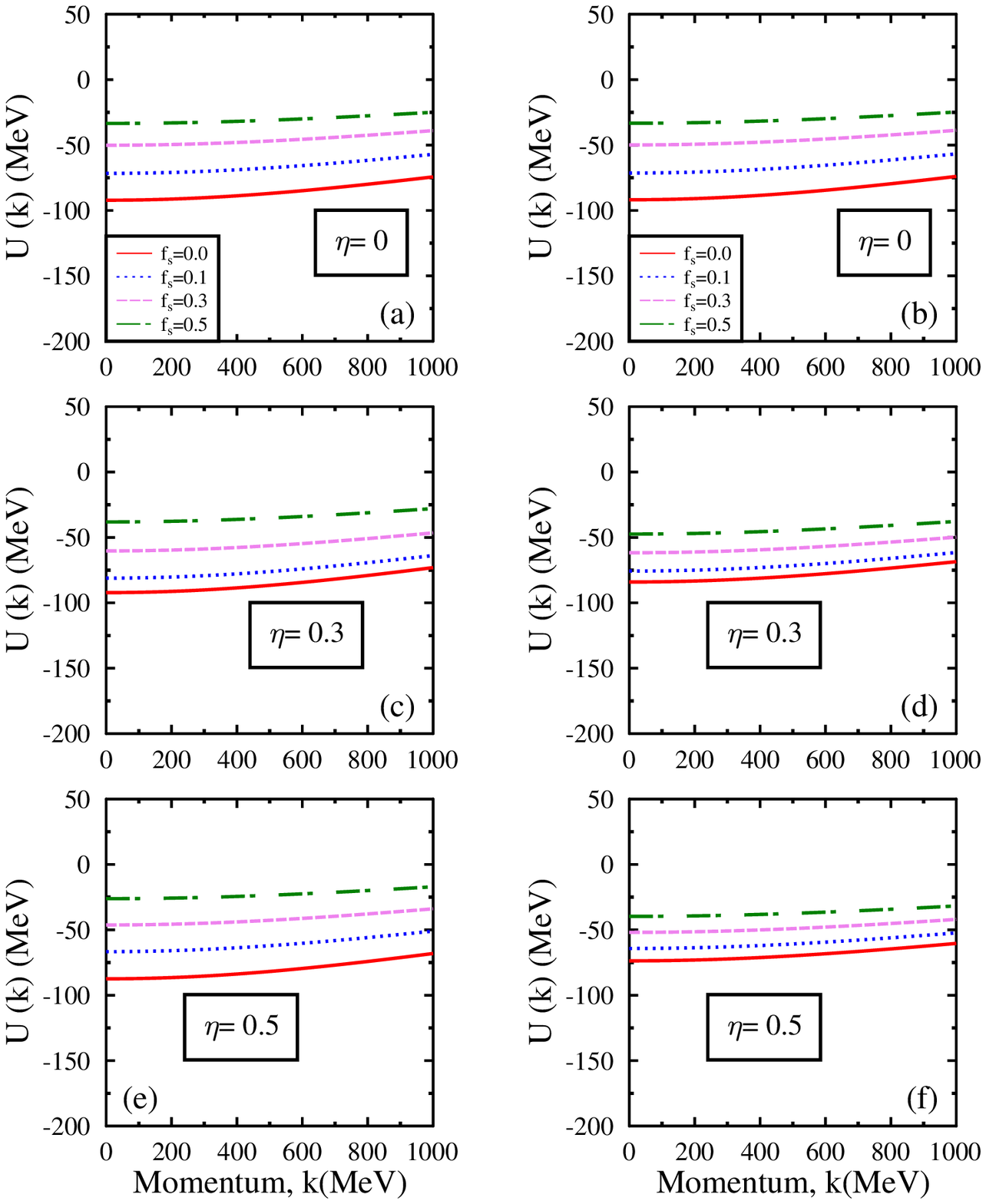} 
\caption{(Color online) The optical potential of $D^{-}$ meson 
(a,c and e) and of $\bar{D^{0}}$ meson (b,d and f), 
are plotted as functions of momentum for $\rho_{B} = 2\rho_0$, 
for different values of the strangeness fractions 
($f_{s} = 0, 0.1, 0.3, 0.5$) and for a given value 
of isospin asymmetry parameter ($\eta = 0, 0.3$ and $0.5$).} 
\label{fig.8}
\end{figure}

\begin{figure}
\includegraphics[width=16cm,height=16cm]{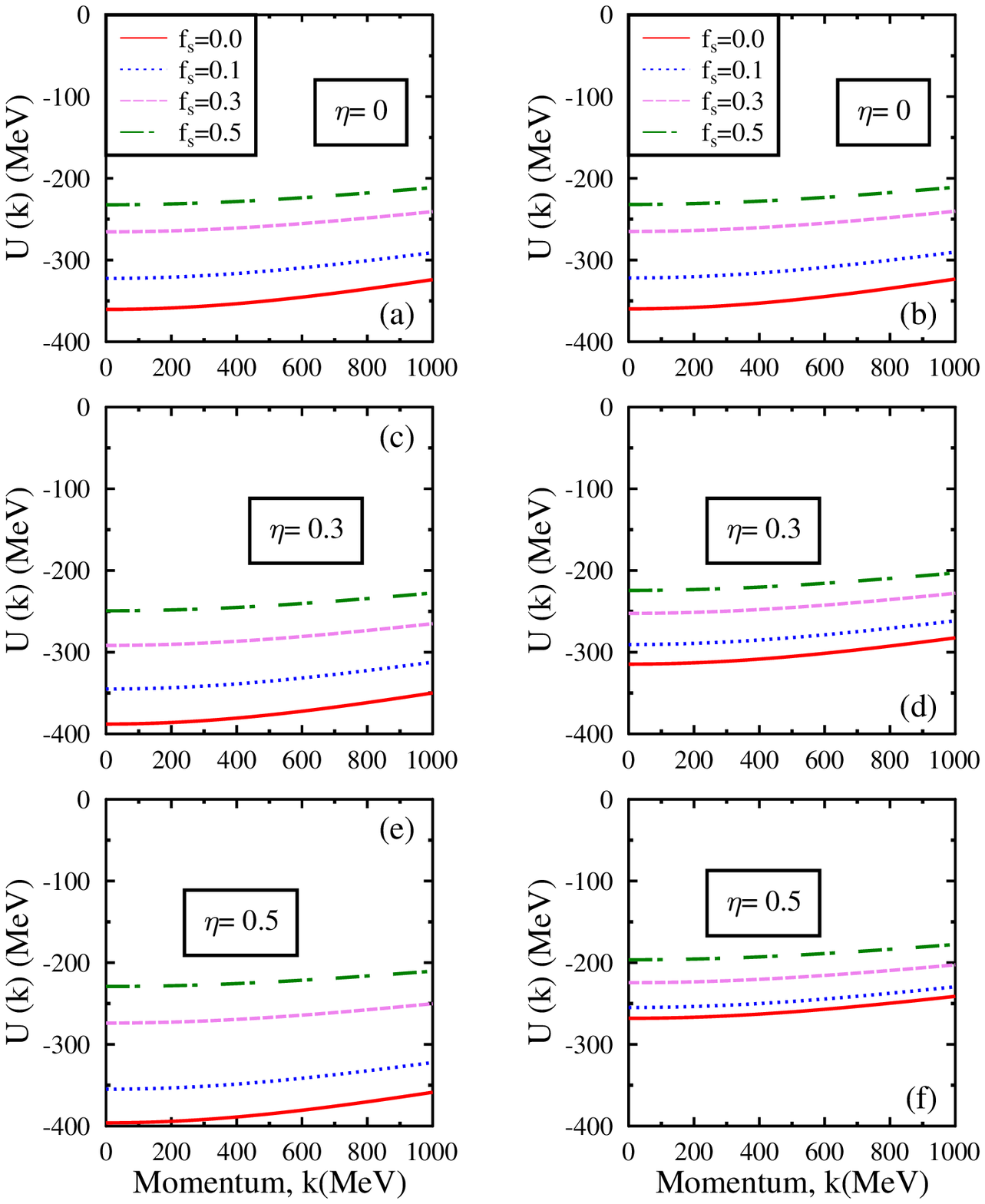} 
\caption{(Color online) The optical potential of $D^{+}$ meson 
(a,c and e) and of $D^{0}$ meson (b,d and f), 
are plotted as functions of momentum for $\rho_{B} = 4\rho_0$, 
for different values of the strangeness fractions 
($f_{s} = 0, 0.1, 0.3, 0.5$) and for a given value 
of isospin asymmetry parameter ($\eta = 0, 0.3$ and $0.5$).} 
\label{fig.9}
\end{figure}

\begin{figure}
\includegraphics[width=16cm,height=16cm]{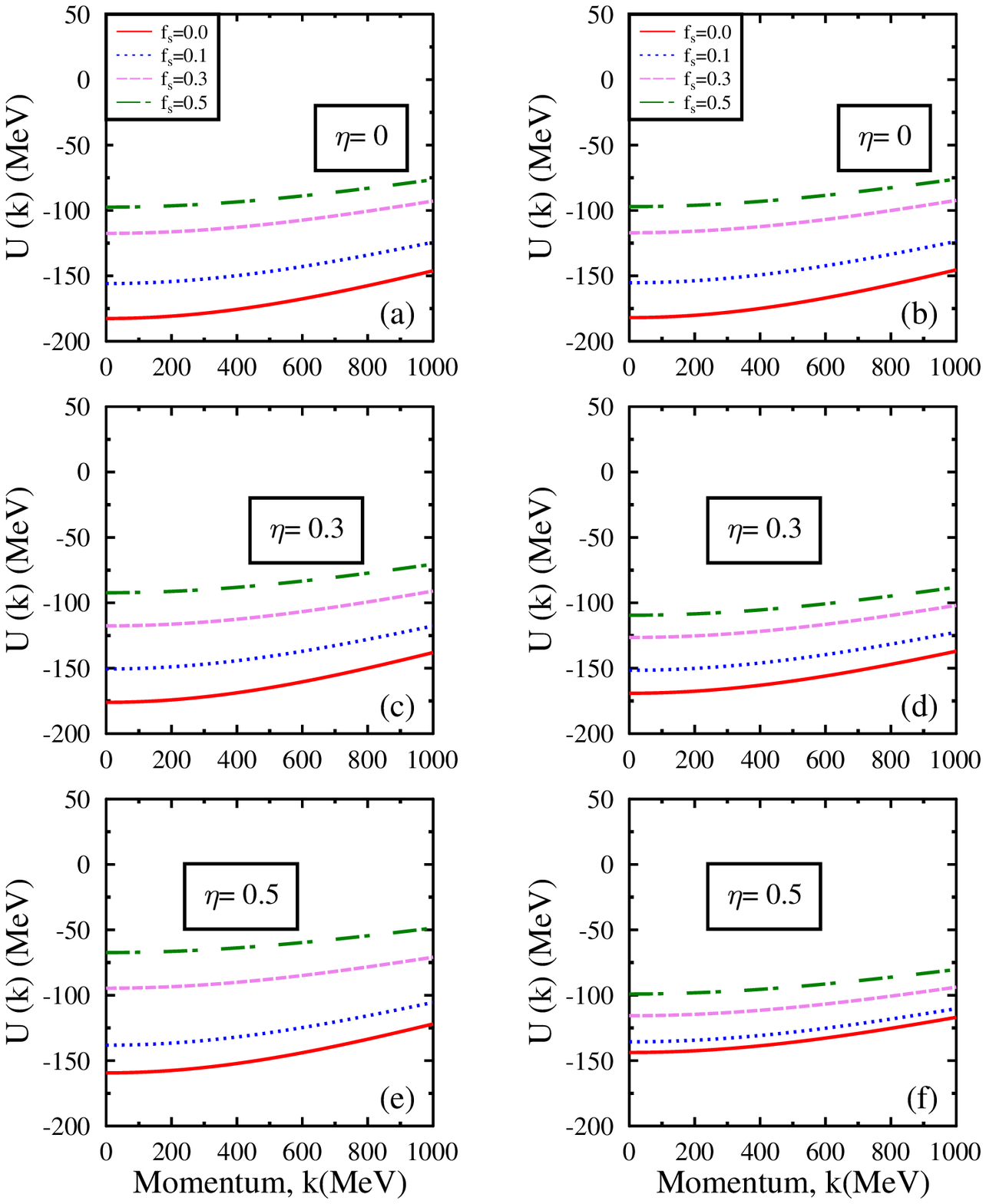} 
\caption{(Color online) The optical potential of $D^{-}$ meson 
(a,c and e) and of $\bar{D^{0}}$ meson (b,d and f), 
are plotted as functions of momentum for $\rho_{B} = 4\rho_0$, 
for different values of the strangeness fractions 
($f_{s} = 0, 0.1, 0.3, 0.5$) and for a given value 
of isospin asymmetry parameter ($\eta = 0, 0.3$ and $0.5$).} 
\label{fig.10}
\end{figure}

The mass modification of the $D$-meson at finite density has been studied 
in the QCD sum rule approach. Borel-transformed QCD sum rules were used 
to describe the interactions of $D$ mesons with nucleons by taking into 
account all the lowest dimension-4 operators in the operator product 
expansion (OPE) and the mass shift at the nuclear matter saturation 
density was found to be around $-50$ MeV \cite{arata}. In the quark meson 
coupling (QMC) model, where the $D(\bar D)$ mesons are assumed to be
bound states of a light quark (antiquark) and a charm antiquark (quark),
interacting through exchange of scalar and vector mesons,
the mass shift of the $D$ meson was calculated to be around 
$-60$ MeV \cite{qmc}. In the present calculations, we observe a drop
in the masses of the $D$ as well as $\bar D$ mesons. However,
the mass drop is seen to be less in the presence of hyperons
in the medium, as can be observed from figures \ref{fig.3} and \ref{fig.4}.
The drop in the masses of the $D$ mesons is also supported by the
recent calculations in the coupled channel approach based
on heavy quark symmetry \cite{garcia1}, which may lead to the 
possibility of the formation of D-mesic nuclei.  
In figures \ref{fig.5} and \ref{fig.6}, we show the optical potentials 
for the $D$ and $\bar{D}$ mesons as functions of momentum,
at the nuclear matter saturation density, and in figures \ref{fig.7},
\ref{fig.8}, \ref{fig.9} and \ref{fig.10}, the optical potentials
for densities of $2\rho_0$ and $4\rho_0$, are plotted. These are 
illustrated for different values of the isospin asymmetry parameter.
The effects of isospin asymmetry and strangeness fraction of the medium 
on the in-medium masses of $D$ and $\bar{D}$ mesons 
are in turn reflected in their optical potentials.
From the plots we observe that the optical potentials of $D$ and $\bar{D}$
mesons do not depend significantly on the momentum of the mesons at 
low densities, but as we move to high densities e.g. to $\rho_{B} 
= 4\rho_{0}$, there does seem to be appreciable momentum dependence. 
It is observed that for a given baryon density, with increase of 
strangeness fraction, $f_{s}$, of the medium, there is an increase in
the magnitude of the optical potential of $D^{+}$ mesons as compared 
to the case of the $D^{0}$ mesons, in isospin asymmetric hyperonic matter.
This is a reflection of the fact that the mass of $D^{+}$ meson has 
a larger drop with increase in the strangeness fraction of the medium,
as compared to  the mass drop of $D^{0}$ meson.
For a given density and strangeness fraction,
the drop in the optical potential of $D^{+}$ meson is increased,
whereas, there is seen to be a decrease in the drop of the 
$D^0$ meson as we increase the isospin asymmetry in the medium. 
This is due to the interaction with the scalar isovector
$\delta$ meson, which has opposite signs for the $D^+$ and $D^0$ mesons,
in the isospin asymmetric hadronic matter, as can been seen from 
equation (\ref{selfd}). The isospin asymmetric contribution
of the Weinberg-Tomozawa term also leads to an increase (decrease)
in the mass drop of the $D^+(D^0)$ meson in the asymmetric hadronic
matter. On the other hand, the dependence
of $D^-$ and $\bar {D^0}$ meson masses on the isospin asymmetry is
seen to be larger for the ${D^-}$ mesons as compared to
the $\bar {D^0}$ meson masses, as can be seen from figure \ref{fig.4}.
These are in turn reflected in their optical potentials,
plotted in figures \ref{fig.6}, \ref{fig.8} and \ref{fig.10},
for densities, $\rho_B=\rho_0, 2\rho_0$ and $4\rho_0$ respectively.
The effect of stranegness is also observed to be more appreciable
for the $D^-$ mesons as compared to the $\bar {D^0}$ mesons.
At high densities, the effects of strangeness as well as
isospin asymmetry are observed to be quite dominant. 
The ratios $D^{+}/D^{-}$ and $D^{0}/\bar{D^{0}}$ could be 
promising experimental tools to verify the effect of strangeness 
fraction of the isospin asymmetric medium on the properties 
of $D$ and $\bar{D}$ mesons in compressed baryon matter asymmetric
heavy ion collision experiments at the FAIR project in the future
facility at GSI.


The medium modifications of the masses of $D$ and $\bar {D}$ mesons can 
lead to an explanation of $J/\psi$ suppression observed by NA50 
collaboration at $158$ GeV/nucleon in the Pb-Pb collisions \cite{blaiz}. 
Due to the drop in the mass of the $D\bar D$ pair in the nuclear medium,
it can become a possibility that the excited states of charmonium 
($\psi^{'}, \chi_{c2}, \chi_{c1}, \chi_{c0}$) can decay to $D\bar{D}$ pairs
\cite{amarind} and hence the production of $J/\psi$ from the decay of 
these excited states can be suppressed. Even at some densities it can 
become a possibility that the $J/\psi$ itself decays to $D\bar{D}$ pairs. 
Thus the medium modifications of the $D$ mesons 
can change the decay widths of the charmonium states \cite{friman}. 

The decay of the charmonium states have been studied in Ref. 
\cite{friman,brat6}. It is seen to depend sensitively on the relative 
momentum in the final state. These excited states might become narrow 
\cite{friman} though the $D$ meson mass is decreased appreciably at 
high densities. It may even vanish at certain momentum corresponding 
to nodes in the wave function \cite{friman}. Though the decay widths 
for these excited states can be modified by their wave functions, the 
partial decay width of $\chi_{c2}$, owing to absence of any nodes, 
can increase monotonically with the drop of the $D^{+}D^{-}$ pair mass 
in the medium. This can give rise to depletion in the  $J/\psi$ yield 
in heavy-ion collisions. The dissociation of the quarkonium states 
($\Psi^{'}$,$\chi_{c}$, $J/\psi$) into $D\bar{D}$ pairs has also been 
studied \cite{wong,digal} by comparing their binding energies with the 
lattice results on the temperature dependence of the heavy-quark 
effective potential \cite{lattice}. 

\section{summary}
We have investigated in the present work, the in-medium masses of the $D$, 
$\bar{D}$ mesons in isospin asymmetric strange hadronic matter. 
The properties of the light hadrons (the nucleons, scalar $\sigma$ meson
and the scalar-isovector $\delta$ meson) in the hadronic medium containing
the nucleons as well as the hyperons -- as studied in  $SU(3)$ chiral 
model -- modify the $D(\bar{D})$ meson properties in the strange hadronic 
matter. The chiral $SU(3)$ model, with parameters fixed from the properties 
of the hadron masses in vacuum and low-energy KN scattering data, 
is extended to SU(4) to derive the interactions of $D(\bar{D})$ mesons 
with the light hadron sector. The mass modifications of $D^{+}$ and $D^{0}$ 
mesons are observed to be more strongly dependent on the isospin-asymmetry 
of the hadronic medium as compared to the medium modifications of the
 masses of $D^{-}$ and $\bar{D^{0}}$ 
mesons. For a given value of density and isospin asymmetry, the strangeness 
of the medium is seen to lead to an increase in the masses of $D$ and 
$\bar{D}$ mesons. 
The effect of strangeness fraction on the in-medium masses of $D$ mesons 
is observed to be larger as compared to those of the $\bar{D}$ mesons.
The mass modification for the $D$ mesons are seen to be similar to the
earlier finite density calculations of D mesons using  QCD sum rule 
approach \cite{qcdsum08,weise} as well as to the results obtained 
using the quark-meson coupling model \cite{qmc}. This is contrary 
to the small mass modifications observed in the coupled 
channel approach \cite{ljhs,mizutani8}. In the present calculations,
we obtain attractive mass shifts for the $D$ as well as the $\bar{D}$ 
mesons, contrary to the repulsive mass shifts for the $\bar D$ mesons
within the coupled channel approach \cite{mizutani8}. The observed 
attractive mass shifts for the $D$ mesons is also supported 
by the recent calculations in coupled channel approach
based on heavy quark symmetry \cite{garcia1}, which can give
rise to the possibility of the $D$-mesic nuclei. 

The ratios $D^{+}/D^{0}$ and $D^{-}/\bar{D^{0}}$ could be 
promising observables to study the effect of strangeness fraction 
of the medium on the properties of $D$ and $\bar{D}$ mesons.
The isospin dependence of 
$D^{+}$  and $D^{0}$ masses is seen to be a dominant medium effect 
at high densities, which might show in their production ($D^{+}/D^{0}$), 
whereas, for the $D^{-}$ and $\bar{D}^{0}$, one sees that, even though 
these have a strong density dependence, their in-medium masses remain 
similar at a given value for the isospin-asymmetry parameter $\eta$. 
The strong density, strangeness and isospin dependence of the $D(\bar{D})$ 
meson optical potentials in asymmetric strange hadronic matter 
can be tested in the asymmetric heavy-ion collision experiments 
at future GSI facility \cite{gsi}. The study of the in-medium modifications 
of $D$ mesons in strange hadronic matter at finite temperatures within 
the $SU(4)$ model will be a possible extension of 
the present investigation. The medium modifications of the charmonium states 
as well as strange charmed mesons in the strange hadronic matter are also 
planned to be investigated.
\acknowledgements 
We thank Arshdeep Singh Mehta for discussions. Financial support from 
Department of Science and Technology, Government 
of India (project no. SR/S2/HEP-21/2006) is gratefully acknowledged. 
One of the authors (AM) is grateful to the Frankfurt Institute for Advanced
Research (FIAS), University of Frankfurt, for warm hospitality and 
acknowledges financial support from Alexander von Humboldt Stiftung 
when this work was initiated.

\end{document}